**Effect of the parameters of bimodal microstructure on the mechanical properties of alumina: A case of sintering regime effects**

M.S. Boldin, E.A. Isupova, E.A. Lantcev, T.S. Pozdova, M.D. Nazmutdinov, D.A. Permin, A.A. Murashov, A.N. Sysoev, A.V. Nokhrin[(*)], V.N. Chuvil'deev

*Lobachevsky University, 603022 Russia, Nizhny Novgorod*

e-mail: boldin@nifti.unn.ru

**Abstract**

The effect of sintering regimes on the density, microstructure parameters, and mechanical properties of $Al_2O_3$ and $Al_2O_3$ + 0.25% MgO ceramics has been investigated. The ceramics were sintered in three regimes: Regime I – heating at a constant rate (2.5, 5, 10, 20 °C·min$^{-1}$) up to the temperature T = 1650°C; Regime II – heating with a varied heating rate up to 1565°C with the duration corresponding to sintering at the heating rate of 10 °C·min$^{-1}$ in Regime I followed by a three-fold decrease in the shrinkage rate; Regime III – two-stage sintering: heating according to Regime II up to the temperature $T_1$ = 1550 °C, then lowering the temperature down to $T_2$ = 1300-1500 °C and holding for 3 h at the $T_2$. The sintering regimes were chosen so that the ceramics had the relative density of 97-99% and a bimodal distribution of the microstructure parameters. The $Al_2O_3$ and $Al_2O_3$ + 0.25% MgO ceramics obtained in Regimes I–III had a microstructure with abnormally large grains in a fine-grained matrix. The sizes and volume fractions of the large grains depended on the sintering regime. Most abnormally large grains had elongated shapes that leads to deviations in the crack propagation trajectories from the straight line. The optimal parameters of the bimodal microstructure parameters distribution providing enhanced mechanical properties of the ceramics (hardness, indentation fracture toughness, ultimate strength) have been determined.

[(*)] Corresponding author (Aleksey Nokhrin, nokhrin@nifti.unn.ru)

## 1. Introduction

Ceramics based on alumina have a wide range of applications in mechanical engineering due to the combination of high hardness, wear resistance, and thermal stability [1-9]. The best mechanical properties of the alumina-based ceramics are achieved through the formation of a high-density fine-grained microstructure [5-6]. Reducing the grain sizes in the ceramics from 5 down to 1 μm leads to an increase in the strength 2 times and more [7]. The maximum ultimate strength of the $Al_2O_3$ ceramics was achieved at the average grain sizes of 1-1.5 μm [8]. The maximum wear resistance was observed in the ceramics with the grain sizes of 0.4-0.65 μm [9]. To obtain ultrafine-grained alumina, hot pressing of high-purity nano- or submicron-grade powders with homogeneous particle size distribution is applied commonly.

Prolonged isothermal holding aimed at increasing the density of the ceramics during conventional pressureless sintering or hot pressing of the nano- or submicron-grade powders leads to the formation of the coarse-grained structure and to the worsening of the mechanical properties of alumina [3]. Reducing the grain growth rate can be achieved by introducing the second-phase particles (e.g., MgO, CaO, $ZrO_2$, SiC, TiC, etc.) into the ceramic to hinder the grain boundary migration in $Al_2O_3$ [10-13]. However, changes in the phase composition may limit the range and conditions of applications of the ceramics significantly. For example, increasing the fraction of the $ZrO_2$ particles reduces the hardness and wear resistance of $Al_2O_3$ [14]. The additions of SiC and TiC alter the dominant wear mechanism of $Al_2O_3$ [7].

An alternative promising approach to enhancing the mechanical properties of the ceramics is the formation of a fine-grained microstructure with bimodal grain size distribution [15]. The presence of the bimodal grain size distribution in the ceramics is attributed to a deviation from the optimal sintering conditions as well as to uncontrolled changes in the composition, particle dispersion, or packing density of the powders. Therefore, the process of forming the bimodal grain size distribution is hard to control, and the multimodal grain size distribution in ceramics is considered as a disadvantage often, which points to the non-optimal manufacturing conditions. Earlier, M.F. Ashby suggested that the formation of the bimodal grains size distribution in ceramics would reduce the crack growth rate [16]. Further investigations indicated that ceramics with the bimodal grain size distribution may manifest enhanced mechanical properties [15, 17].

The experimental confirmation of the potential to enhance the strength of aluminum oxide by forming a bimodal structure was first reported in Ref. [18]. However, the scale of increasing the strength of aluminum oxide due to the formation of a bimodal structure is small [18]. The small scale of the effect of the bimodal structure on the strength of aluminum oxide is primarily due to the fact that, in most cases, the strength and crack resistance of grain boundaries often exceeds the characteristics of the crystal lattice. The presence of large grains in the structure can cause the crack propagation trajectory to deviate from a straight line, thereby increasing the energy dissipation associated with crack growth. For this to occur, it is necessary that the crack propagates primarily along the grain boundaries. If the boundaries have high strength/crack resistance, then crack propagation will be primarily transcrystalline (Fig. 1a), and the strength will weakly depend on the parameters of the grain structure of the material. In this case, the grain boundaries will act only as barriers to crack propagation, while the appearance of abnormally large grains will reduce the strength (Fig. 1b). To increase the strength of materials with a bimodal structure, it is necessary that the boundaries of abnormally large grains have lower strength/crack resistance compared to the crystal lattice of the grain. In this case, the crack will primarily propagate along the grain boundary. A decrease in the strength of the boundaries of abnormally large grains will lead to a significant deviation of the crack trajectory from a straight line (Fig. 1c) and, as a result, to an increase in the strength of ceramics. Indirectly, the effectiveness of this approach has been confirmed in studies of ceramic matrix composites reinforced with extended fibers [19].

To increase the strength of ceramics with a bimodal structure, it is necessary that the size of large grains significantly exceed the average grain size of the matrix. To decrease the grain size of the matrix, it is possible to use small concentrations of second-phase particles, which reduce the grain growth rate, or optimize the sintering regimes of ceramics.

It is well established that low concentrations of MgO result in a decrease in grain size and promote a more uniform microstructure, along with an increased density of sintered ceramics [20, 21]. In Ref. [22], the dependence of the compaction rate of $Al_2O_3$ during hot pressing on the concentration of MgO was analyzed. It has been shown that an increase in the concentration of MgO to 400 ppm leads to a 3-fold increase in the compaction rate. When the MgO content ranges from 400 to 1000 ppm, the compaction rate remains unchanged. A further increase in the concentration of MgO to 2000 ppm results in a further increase in the compaction rate. The authors also emphasize that the solubility limit of MgO in $Al_2O_3$ at the specified sintering temperature ($T_{sint}$ = 1630 °C) is 250 ppm. In Ref. [23], the effectiveness of the MgO additive for reducing the grain size of aluminum oxide during Spark Plasma Sintering (SPS) was demonstrated. The reduction effect becomes more pronounced with increasing sintering temperature. The authors also note that the microstructure of a sample containing 300 ppm MgO is more uniform than that of a sample containing 500 ppm MgO. However, there are no abnormally large grains in the structure of the samples containing MgO. In Ref. [24], the optimal concentration of MgO in alumina ceramics produced by SPS was determined. The addition of MgO to $Al_2O_3$ in the amount of 1250 ppm leads to an increase in density (by 0.2 %) and a decrease in grain size by 0.3 μm. A further increase in the concentration of MgO to 5000 ppm leads to a sharp decrease in density and a significant increase in the grain size of $Al_2O_3$.

Optimizing sintering regimes is also an effective approach to improving the mechanical properties of ceramics and a convenient way to control the parameters of the bimodal grain size distribution in alumina. To optimize the microstructure parameters of the ceramics, various methods are applied, such as adjusting the heating rates, sintering temperatures, holding times, and pressure applied as well as optimizing the particle size distribution in the powder, etc. Among the most popular approaches for optimizing the sintering conditions, the following methods can be outlined: Response Surface Methodology (RSM) [25-29], which does not require the construction of a physical model of the sintering process; Master Sintering Curve (MSC) [30, 31], which is based on the dilatometric measurements results and the assumption that only one dominant mechanism of diffusion mass transport occurres during the sintering process; the kinetic analysis [32-34] assuming the construction of a multi-stage model of the process describing each sintering stage with a separate equation. An efficient method for optimizing the structure is also Two-Step Sintering (TSS) [35-42], which involves heating to a temperature corresponding to relative density ~ 80-90% and then cooling by several tens of degrees followed by holding for several hours. When optimizing the sintering conditions, the model-free methods for determining the activation energy may be applied also, for example, Wang-Rajah method [43], Friedman method [34], and Ozawa-Flynn-Wall method [32], which allow determining the parameters of the sintering curves.

The aim of the present study was to investigate the effect of the sintering conditions of the submicron-grade alumina powders on the parameters of the high-density fine-grained microstructure and the mechanical properties of the ceramics with the bimodal grain size distribution. The bimodal grain size distribution in alumina originates from the abnormal grain growth (AGG). The sizes and the volumetric fraction of the abnormally large grains were controlled by choosing the method for sintering the submicron-grade alumina Ca-doped powders. The peculiarities of the synthesis conditions of the $\alpha$-$Al_2O_3$ Ca-doped powders resulted in the emerging of the large elongated grains, the form factor ($F$) of which ($F = L_{AGG} / l_{AGG}$ where $L_{AGG}$ and $l_{AGG}$ are the length and width of the abnormally large grains, respectively) depended on the sintering conditions. A larger value of the form factor $F$ allows the deviation of the crack trajectory from a linear one thereby providing additional dissipation of the elastic energy. The addition of 0.25% MgO was used to accelerate the sintering of alumina via increasing the density of the ceramic, affecting the grain growth rate, and affecting the parameters of the bimodal grain size distribution in the ceramics [44, 45]. To reduce the strength of the boundaries of abnormally large grains, calcium microadditives were used [46]. Furthermore, the addition of calcium stimulated abnormal grain growth. This made it possible to obtain fine-grained ceramics with abnormally large grains, the grain boundary strength and crack resistance of which were lower than those of the $Al_2O_3$ crystal lattice.

## 2. Materials and methods

The present study was focused on the α-$Al_2O_3$ powder (Powder A) and its mixture with 0.25 wt.% MgO (Alfa-Aesar, 100 nm grade) (Powder B).

In the synthesis of the α-$Al_2O_3$ powder, aluminum nitrate crystallohydrate $Al(NO_3)_3 \cdot 9H_2O$ (A grade), ammonium bicarbonate $NH_4HCO_3$ (A grade), concentrated $HNO_3$ (ACS grade), concentrated ammonium hydroxide 25% $NH_4OH$ (ACS grade) and 99.8% isopropanol (ACS grade) were used as the raw materials. The 0.7M water solution of aluminum nitrate and the 2M one of ammonium hydrocarbonate were prepared. The precursors were synthesized by the reverse precipitation method from the solution. The cationic solution containing 0.2 M of Al ions was added dropwise at the rate of 2 ml/min to the precipitant solution containing 0.6 M of ammonium hydrocarbonate with constant stirring. The processes, which took place can be described by the following equation: $4NH_4HCO_3 + Al(NO_3)_3 \cdot 9H_2O \rightarrow NH_4AlCO_3(OH)_2\downarrow + 3NH_4NO_3 + 10H_2O + 3CO_2\uparrow$. Throughout the synthesis, the pH was maintained at 8 by adding ammonium hydroxide to the solution. After adding aluminum nitrate, the resulting suspension was stirred constantly for 30 min. Next, the precipitate obtained was washed in two stages: 3 times in distilled water to remove the traces of ammonium hydrocarbonate and 3 times in water-free isopropanol to remove the residual water. The resulting precipitate was dried at 60°C for 8 h (pressure of 600 bar) and then calcined in air in a muffle furnace at the 1150°C (1 h) to obtain α-$Al_2O_3$. As a result of the final annealing, the precursor decomposed to form alumina. The synthesis scheme for the α-$Al_2O_3$ powders is presented in Fig. 2. The synthesized powder contained 86-130 ppm of calcium, which promoted the formation of large elongated grains upon heating [47]. This concentration of Ca corresponds to 0.0086-0.0130 wt.% CaO. The chemical analysis of the powders was performed using the ICP method.

The specific surface area $S_{BET}$ of the precursor and alumina powders was evaluated using the Brunauer–Emmett–Teller (BET) method using Sorbi®-M instrument (META, Russia). Assuming the spherical shape of the powder particles obtained, the average particle diameter $D_{BET}$ was calculated as $6/(\rho \cdot S_{BET})$ where $\rho = 3.98$ g/cm³ is the theoretical density (3.98 g/cm³ for α-$Al_2O_3$ [48]; 2.03 g/cm³ for ammonium aluminum carbonate hydrogen (AACH) [49]).

The powder compacts with a diameter of 5 mm and a height of 2.5 mm were prepared in a steel die by pressing at a pressure of 200 MPa. The sintering kinetics of the pressed powder compacts was studied using Linseis L75 dilatometer with LVDT detector. Heating was performed in air applying constant uniaxial mechanical stress. The temperature was measured with a Pt-Rh type B thermocouple installed near the sample. The temperature measurement uncertainty was ±5°C. The load applied during sintering was 500 ± 15 mN, corresponding to the uniaxial stress of 0.05 MPa. During heating, the effective shrinkage (*L*) and the shrinkage rate (*S*) of the powders were recorded as a function of temperature and heating time. The uncertainty of the shrinkage measurement was ±1-5% of the maximum shrinkage value. Standard dilatometer software was used to account for the contribution of thermal expansion of the measuring system to the effective shrinkage of the powder.

Three sintering regimes were used to sinter the ceramics. These regimes were selected on the basis of the analysis of previous studies:

Regime I - heating at a constant rate ($V_h$) from 2.5 to 20 °C·min$^{-1}$ up to the temperature of 1650°C; no isothermal holding was applied ($t_s = 0$). Regime I provided an initial estimate of the sintering rate for the ceramics. The aim of this series of experiments was to investigate the effect of heating rate on the compaction rate of the powder compacts, ceramic density, sizes, and volumetric fraction of the abnormally large alumina grains. The ceramics obtained in Regime I are characterized usually by a large volumetric fraction of the abnormally large grains. Local variations in the density of the powder compact lead to non-uniform compaction during the sintering process causing the uncontrolled abnormal grain growth while reducing the compaction rate allows increasing the volume uniformity of the sintered material [3].

Regime II - multistage heating ($V_{h1}$ = 20 °C·min$^{-1}$ to $T_1$ = 1100 ºC, $V_{h2}$ = 10 °C·min$^{-1}$ to $T_2$ = 1240 ºC, $V_{h3}$ = 4 °C·min$^{-1}$ to $T_3$ = 1385 ºC, and heating with the rate of 3 °C·min$^{-1}$ up to $T_4$ = 1565 ºC). The duration of sintering in Regime II was equal to the one of sintering in Regime I with $V_h$ = 10 °C·min$^{-1}$. The temperature-time heating conditions were calculated using the NETZSCH Kinetics Neo software, based on the analysis of shrinkage rates S(T) for Regime I. The heating stage was divided into sub-stages ($T_1$, $T_2$, $T_3$, $T_4$) using the NETZSCH Kinetics Neo software, allowing Regime II to reduce the maximum shrinkage rate by 3 times compared to sintering Regime I (at $V_h$ = 10 °C·min$^{-1}$). Additionally, Regime II was used to increase the uniformity of volume shrinkage of the ceramic sample.

For controlling the abnormal grain growth at the final stage of sintering, two-stage Regime III was applied: heating to the temperature $T_1$ = 1550 °C according to Regime II (for forming a homogeneous microstructure with a density $\rho_{rel}$ ~ 90%) followed by a 3-hour isothermal holding at reduced temperature $T_2$ = 1300-1500 °C. The regime was used to implement the two-step sintering approach [50].

The ceramics were cooled down in an uncontrolled regime together with the furnace. The regimes I-III were chosen on the basis of analysis of previous studies to achieve the relative density of the ceramics 97-99%.

The X-ray diffraction (XRD) investigation were carried out using XRD-7000 diffractometer (Shimadzu, Japan) using $CuK_\alpha$ radiation (λ = 1.54 Å). The measurements were performed using the Bragg-Brentano geometry in the angle range 2θ = 20–90° with the step of Δ2θ = 0.04° and the dwell time of 1 s. There was no slit in front of the detector. The phase analysis was conducted using the DIFFRAC.EVA software (Bruker, Germany) utilizing the PDF-2 database (ICDD, 2012).

To determine the temperatures of the phase transformations in the alumina precursor, thermogravimetric/differential scanning calorimetry (TG/DSC) method was applied. The investigations were performed using synchronous Netzsch00® STA 449F1 thermal analyzer in the temperature range from 60 to 1200 °C in Ar flow. The heating rate was 10°C/min.

The powder and ceramics were investigated by scanning electron microscopy (SEM) using Hitachi® Regulus™ SU8100 and JEOL® JSM-6490 microscopes. On the fracture surfaces of the ceramics with the bimodal grain size distributions, the average grain sizes of the fine-grained matrix ($d_{av}$), the lengths ($L_{AGG}$), widths ($l_{AGG}$), and volumetric fractions ($f_v$) of the abnormally large grains were determined. GoodGrains software was used to determine the grain sizes in the fine-grained matrices and the geometrical parameters of the abnormally large grains. The grain sizes were measured using the chord method. The form factors $F$ (the degrees of elongation) of the abnormally large grains were calculated using the formula $F = L_{AGG}/l_{AGG}$. The effective areas of the abnormally large grains were calculated using the formula $S_{AGG} = L_{AGG}l_{AGG}$. The size distributions of the particles were determined using dynamic light scattering (DLS) using NanoBrook 90 PlusZeta instrument. Before the DLS investigations, the powders were subjected to ultrasonic treatment for 5 min.

The initial density of the compacts ($\rho_0$) was determined by weighing and measuring the sample dimensions with a caliper (graduation $\delta = 0.1$ mm). The densities of the sintered specimens ($\rho$) were measured by hydrostatic weighing using Sartorius® CPA 225D balance. When calculating the relative density ($\rho_{rel} = \rho / \rho_{th}$), the theoretical density of pure alumina was taken as $\rho_{th} = 3.96$ g/cm$^3$. The initial relative density of the powder compacts was approximately 40%.

The microhardness (Hv) was measured using Qness® A60+ hardness tester at the load P = 2 kg. The fracture toughness ($K_{IC}$) during the indentation was calculated using Palmquist method based on measuring the length of the longest radial crack:

$$K_{IC} = 0.016 \frac{P}{c^{3/2}} \sqrt{E/H_v} \tag{1}$$

where $c$ is the average distance from the center of the imprint to the end of the longest radial crack [m] and E = 370 GPa is Young modulus of the material. The average uncertainties of the $H_v$ and $K_{IC}$ measurements were ± 1 GPa and ± 0.3 MPa·m$^{1/2}$, respectively.

The ultimate strength of the ceramics was investigated using the B3B (Ball on Three Balls) loading method [51-55]. The B3B test is one of the most effective methods to evaluate the mechanical properties of ceramics used in conditions of a multiaxial stress-strain state. On the one hand, this makes it difficult to analyze the results, but it allows for a more reliable assessment of the effect of the parameters of a bimodal structure on the behavior of the material under conditions that close to operational conditions. This study represents the first investigation of B3B strength in ceramics with a bimodal structure. To determine the ultimate strength ($\sigma_b$), the samples with the diameter of 12 mm and the thickness of 0.5 mm were manufactured. The samples for the mechanical tests were made in Nabertherm® LHT 0418 furnace according to Regimes I-III. The identity of samples manufactured in the Nabertherm® LHT 0418 furnace and by sintering in the dilatometer was controlled by measuring the density, microhardness, and examining the microstructure. A universal breaking machine 2167 P-5 was used for the B3B strength testing. The loading speed was 1 mm/min, the diameters of the supporting steel balls were 5.53 mm, the radius of the circumscribed circle around the triangle formed by the centers of the steel balls was $R_a$ = 5.35 mm. The testing method was described in [56]. Three samples of each microstructure type or more were tested. The average uncertainty of determining the ultimate strength was 50 MPa.

## 3. Results

### 3.1 Synthesis and characterization of the α-$Al_2O_3$ powders

Fig. 3 shows the XRD curves of the precursor and synthesized α-$Al_2O_3$ powder. Only the $NH_4Al(OH)_2CO_3$ phase (PDF-2 № 01-071-1314) was detected in the sample. Significant broadening of the XRD peaks indicates indirectly the presence of the particles with the sizes < 50 nm. The results of the XRD investigations have shown the powders to consist of α-$Al_2O_3$ (PDF #00-046-1212) and small amounts of the θ-$Al_2O_3$ (PDF #01-079-1559) and γ-$Al_2O_3$ (PDF #01-076-4179) impurity phases were detected also. The XRD peaks of the θ-$Al_2O_3$ and γ-$Al_2O_3$ phases are characterized by significant broadening as compared to the peaks of the α-$Al_2O_3$ phase. This indicates indirectly that the particle sizes of the impurity phases were significantly smaller than the ones of the particles in the main phase. There were no halos in the diffraction patterns, which are characteristic of amorphous materials. The MgO powders were the single-phase ones and had the cubic structure (PDF #04-010-4039).

Fig. 4 shows the DSC and TG curves obtained during the continuous heating of the precursor powder in the Ar ambient. The DSC curve manifested two endothermic effects ($T_1$ = 126 °C, $T_2$ = 224 °C), which were attributed to the removal of the absorbed moisture and to the decomposition of the precursor into amorphous alumina, respectively. Similar effects were observed in [57] in the DTA curve for the decomposition of the AACH precursor. The absence of an exothermic peak at 270–280 °C (see [57]) was associated with the fact that the precursor had a crystalline structure initially. The thermogram of the sample is characterized by a tendency to the mass loss with increasing temperature. The sample's mass reached a saturated value in the temperature range from 800 to 1000 °C.

The specific surface area of the synthesized alumina powders was $S_{BET}$ ≈ 33 $m^2/g$ corresponding to the average particle sizes $D_{BET}$ ≈ 45 nm. The specific surface area of the precursor powders was 737 $m^2/g$ ($D_{BET}$ ≈ 4 nm). These results indicate the coagulation and agglomeration of alumina particles to occur during synthesis at 1150°C.

The SEM images of the initial powders are shown in Fig. 5. One can see the $\alpha$-$Al_2O_3$ and MgO powders to consist of the nanoparticles of ~100 nm in sizes. The alumina particle size distribution is presented in Fig. 6. From Fig. 6, one can see that the synthesized nanopowder was agglomerated; the sizes of the agglomerates varied from 200 to 500 nm. The largest fraction of agglomerates had the sizes of ~400 nm (Fig. 3). In our opinion, the high specific surface area $S_{BET}$ and small effective particle sizes $D_{BET}$ are primarily due to the presence of a significant number of the nanometer-sized $\alpha$-$Al_2O_3$ particles in the synthesized powder.

3.2 Sintering in Regime I

The temperature dependencies of the shrinkage rate $S(T)$ are presented in Fig. 7. The $S(T)$ dependencies had a multi-stage character, primarily due to the differences in the sintering rate of the nanoparticles in the agglomerates and the one of the agglomerates between each other. This assumption was verified by studying the effect of the preliminary pressing pressure on the character of the $S(T)$ dependencies.

In the $S(T)$ dependencies of powder A at the heating rate $V_h$ = 20 °C/min (Fig. 7a), two maxima ($S_{max}$) were identified: one at the temperatures around ~1440°C ($S_{max(1)} = 2.5\cdot10^{-5}$ %·s$^{-1}$) and another one ~1540°C ($S_{max(2)} = 3.4\cdot10^{-5}$ %·s$^{-1}$). Decreasing the heating rate led to a reduction in the shrinkage rate $S$, an increase in the width of the maxima, and their shift towards lower temperatures. The magnitude of the low-temperature maximum at 1100-1150°C and $V_h \leq 10$ °C·min$^{-1}$ was very small, and it was not considered in further analysis of the $S(T)$ curves. Thus, for the heating rate 10 °C·min$^{-1}$, the peaks in the temperature curve $S(T)$ were observed at the temperatures $T_{max(1)}$ = 1420°C ($S_{max(1)} = 1.3\cdot10^{-5}$ %·s$^{-1}$) and $T_{max(2)}$ = 1530°C ($S_{max(2)} = 1.7\cdot10^{-5}$ %·s$^{-1}$), and for a the heating rate of 2.5 °C·min$^{-1}$ – at the temperatures $T_{max(1)}$ = 1370°C ($S_{max(1)} = 0.33\cdot10^{-5}$ %·s$^{-1}$) and $T_{max(2)}$ = 1510°C ($S_{max(2)} = 0.45\cdot10^{-5}$ %·s$^{-1}$). A summary of the results obtained is presented in Table 1.

The character of the curves $S(T)$ for powder B was similar to the one for powder A. However, the addition of 0.25% MgO affected the maximum shrinkage rate with respect to the temperatures, which the maxima were observed at in the $S(T)$ curves. When adding 0.25% MgO to alumina (powder B), the value of $S_{max(1)}$ on the temperature curve $S(T)$ at the heating rate 20 °C·min$^{-1}$ almost did not change and amounts to $2.4\cdot10^{-5}$ %·s$^{-1}$; the magnitude of the second maximum increased up to $S_{max(2)} = 4.1\cdot10^{-5}$ %·s$^{-1}$. The maximum shifted towards lower temperatures by ~30°C (Fig. 7b). Similar effects of MgO on the sintering kinetics of the $Al_2O_3$ powders were observed at other heating rates.

The heating rate did not affect significantly the densities of the ceramics obtained. The average value of the relative density of the ceramics A was 96.3 ± 1.3%. The addition of MgO led to an increase in the density of ceramics B up to ~97.5 ± 1.3%.

Fig. 8 shows characteristic microstructure images of the ceramics obtained in Regime I. One can see in Fig. 8 the microstructure of the samples obtained by sintering at constant heating rate 10 °C·min$^{-1}$ to be non-uniform, with fine grains with the average sizes of 0.8-0.9 μm and large elongated grains. The micron-sized pores were observed inside the large grains. The lengths ($L_{AGG}$) and widths ($l_{AGG}$) of the large grains in ceramic A were 30 ± 14 μm and 8 ± 5 μm, respectively. In ceramic B, the lengths of the large grains were 16 ± 9 μm and their widths were 8 ± 4 μm. Thus, the form factors *F* of the large grains for ceramics A and B obtained in Regime I at 10 °C·min$^{-1}$ were ≈ 4 and ≈ 2, respectively. The volume fractions of the abnormally large grains ($f_v$) were 51% and 12% in ceramics A and B, respectively. Therefore, the addition of MgO reduces the volume fraction of the abnormally large grains, the elongation degree, and the sizes of the large grains.

Increasing the heating rate up to 20 °C·min$^{-1}$ did not affect the microstructure of the ceramics significant while reducing $V_h$ down to 2.5 °C·min$^{-1}$ led to an increase in the volume fraction of the large grains up to ~90% and to slight increase in their lengths and widths. The form factors of the grains at 2.5 °C·min$^{-1}$ remained almost unchanged (≈ 4.5 for ceramic A and ≈ 2 for ceramic B). The results of investigations of the ceramics are summarized in Table 2, which also includes the values of the effective area of the abnormally large grains $S_{AGG}$.

The addition of MgO led to an increase in the hardness and fracture toughness of alumina. The hardness values for ceramics A and B were in the range of 13.1-14.0 and 16.2-19.5 GPa, respectively. The fracture toughness $K_{IC}$ ranged from 2.0-2.6 and 4.5-4.9 MPa·m$^{1/2}$, respectively. The ultimate strength of Ceramic A sintered with the heating rate of 10 °C/min was 300 ± 40 MPa. For Ceramic B, $\sigma_b$ = 520 ± 50 MPa.

### 3.3 Sintering in Regime II

The temperature curves *S*(*T*) for pure alumina and $Al_2O_3$ + 0.25% MgO ceramics are presented in Fig. 9. The presence of four local maxima ($T_1$, $T_2$, $T_3$, and $T_4$) on the *S*(*T*) curve was associated with the changes in the heating rate $V_h$ at these temperatures (see Materials and methods section). The average sintering rate over the entire temperature range of the sintering process was approximately 1/3 of the maximum one in Regime I (for 10 °C·min$^{-1}$). The sintering rate of powder A [(0.35 – 0.55)·10$^{-5}$ %·s$^{-1}$] was slightly lower than the one of powder B ((0.35–0.72) ·10$^{-5}$ %·s$^{-1}$) (Fig. 9).

The relative densities of ceramics A and B were 94.5 ± 0.9% and 97.1 ± 0.2%, respectively. Therefore, the density of the ceramics sintered in Regime II was close to the one of ceramics sintered in Regime I (Table 2).

The microstructure of the ceramics is shown in Fig. 10. One can see the microstructure of the ceramics sintered in Regime II to be more uniform compared to the one of ceramics sintered in Regime I. The major volume fractions of the ceramics A and B were characterized by fine grained microstructure with equiaxial grains; the average grain sizes $d_{av}$ = 0.7 μm. The large grains in ceramic A had the length $L_{AGG}$ = 29 ± 21 μm and the width $l_{AGG}$ = 5 ± 2 μm. The large grains in ceramic B had smaller lengths $L_{AGG}$ = 19 ± 11 μm but larger widths $l_{AGG}$ = 8 ± 4 μm. This indicates a higher degree of roundness and a lower form factor *F* for ceramic B than the ones in ceramic A. The volumetric fractions of large grains in ceramics A and B were comparable to each other: $f_v$ ~ 5%.

The hardness of ceramic A was 15.7 GPa, the fracture toughness was 3.0 MPa·m$^{1/2}$. The addition of MgO (ceramic B) resulted in an increase in the hardness up to 18.4 GPa and in the fracture toughness up to 5.0 MPa·m$^{1/2}$. The cracks exhibited predominantly linear propagation within the fine-grained matrix. However, in the presence of large grains, the crack trajectories changed tending to bypass the large grains. The ultimate strength of Ceramic A sintered in Regime II was 255 ± 30 MPa, for Ceramic B, $\sigma_b$ = 330 ± 40 MPa. The comparison of the properties of ceramics obtained in Regime II shows the ceramics with the addition of MgO to have higher density, hardness, and fracture toughness. For this reason, the next series of experiments aimed at obtaining the bimodal grain size distribution was carried out using powder B.

3.4 Sintering in Regime III

The microstructure of the ceramics obtained in Regime III is shown in Fig. 11. From Fig. 11, one can see the microstructure of the ceramics obtained at the temperatures $T_2$ = 1300 and 1350°C to have a unimodal grain size distribution and to consists mainly of the grains with the sizes of 0.7 μm.

Also, some isolated large grains with the lengths $L_{AGG}$ = 13 ± 5 μm and the widths $l_{AGG}$ = 5 ± 2 μm were observed. The volumetric fraction of the large grains did not exceed 2-5%.

In the ceramic obtained at the temperature $T_2$ = 1400 °C, the elongated large grains with the lengths $L_{AGG}$ = 23 ± 12 μm, the widths $l_{AGG}$ = 10 ± 4 μm, and the volumetric fraction of 30% were observed against the fine-grained microstructure with the average grain sizes of 0.7 μm. Increasing the isothermal holding temperature up to 1450°C led to an increase in the sizes and volumetric fraction of the large grains: $L_{AGG}$ = 35 ± 21 μm, $l_{AGG}$ = 14 ± 9 μm, and $f_v$ = 60% while maintaining the average sizes of the fine particles (0.7 μm). The characteristic grain size distribution histogram is presented in Fig. 11f.

In ceramic B sintered at $T_2$ = 1500°C, the clusters of fine grains of 0.8 μm in sizes were observed along with predominantly elongated large grains with the lengths $L_{AGG}$ = 55 μm and the widths $l_{AGG}$ ~ 29 μm. The volume fraction of the large grains in this ceramic reached 78%. The results of the microstructure investigations of the ceramics are summarized in Table 3.

The average relative density of ceramics B obtained in Regime III was 97.8 ± 2.0%. The confidence intervals of the density values of the samples obtained at different temperatures $T_2$ overlapped.

Ceramic B sintered in Regime III at 1350°C exhibited relatively high hardness Hv = 19.0 ± 4.1 GPa and fracture toughness $K_{IC}$ = 5.4 MPa·$m^{1/2}$. The results of the metallographic investigations on the surfaces of the ceramics after hardness measurements revealed an increase in the volumetric fraction of the abnormally large grains that affected the crack propagation trajectories significantly. Note that in the ceramics with increased volumetric fraction of the abnormally large grains, a substantial deviation of the crack propagation trajectories from the straight-line ones was observed.

## 4. Discussion

### 4.1 Analysis of sintering kinetics

For the analysis of the sintering kinetics of the alumina powders, it is most convenient to use the $L(T)$ and $S(T)$ dependencies obtained for Regime I (Fig. 7). From Fig. 7, one can see the $S(T)$ dependencies to manifest typical two-stage character. Let us analyze the possible sintering mechanisms at each compaction stage. Prior to the analysis, it should be noted again that the α-$Al_2O_3$ powder contains an increased concentration of Ca, which provokes anomalous grain growth in alumina.

To analyze the compaction of powders at the initial stage of intensive densification, it is convenient to use the Young-Cutler model [58]. This model describes the sintering process of spherical particles under the simultaneous occurrence of volume diffusion, grain boundary diffusion, and viscous flow (creep) of the material. The effective sintering activation energy ($mQ_{s1}$, [$kT_m$]) of the ceramic in the Young-Cutler model can be determined from the slope of the dependence of relative shrinkage ε on the heating temperature $T$ in the $\ln(T\partial\varepsilon/\partial T) - T_m/T$ where $T_m$ = 2317 K is the melting point of alumina, $m$ is a numerical factor dependent on the sintering mechanism ($m$ = ½ for the volume diffusion, $m$ = 1/3 for the grain boundary diffusion, and $m$ = 1 for the viscous flow (creep) of the material). The efficiency of applying the Young-Cutler model to describe the sintering process of alumina powders in the continuous heating regime was demonstrated earlier in [59-62].

To analyze the compaction kinetics of the powders at the second sintering stage where a decrease in the powder shrinkage rate was observed (Fig. 5), it is convenient to use the model of diffusion-controlled pore annihilation near the grain boundaries [63-65]. The sintering activation energy ($Q_{s2}$, [$kT_m$]) of the ceramic within this model can be determined from the slope of the dependence of the compaction $\rho/\rho_{th}$ on the heating temperature in double logarithmic axes $\ln\{\ln[(\rho/\rho_{th})/(1-\rho/\rho_{th})]\} - T_m/T$. The procedure of calculating the temperature dependence of compaction $\rho/\rho_{th}(T)$ from the one of shrinkage $L(T)$ is described in [65].

As an example, Fig. 12 shows the dependencies $\ln(T\partial\varepsilon/\partial T) - T_m/T$ and $\ln\{\ln[(\rho/\rho_{th})/(1-\rho/\rho_{th})]\} - T_m/T$ for the $Al_2O_3$ powders heated at the rates of 2.5 and 20 °C·min$^{-1}$. From Fig. 12, one can see the dependencies $\ln(T\partial\varepsilon/\partial T) - T_m/T$ to have a two-stage character. The slopes of the $\ln(T\partial\varepsilon/\partial T) - T_m/T$ dependencies in the low-temperature range [$mQ_{s1}$ = 15.0–22.9 $kT_m$] is much larger than the effective activation energy for sintering at higher temperatures [5–6 $kT_m$]. The temperatures, which the decreases in the effective sintering activation energy were observed at were ≈1100–1300°C for $V_h$ = 2.5 °C·min$^{-1}$ and ≈1140–1170 °C for $V_h$ = 20 °C·min$^{-1}$. According to [66], the chemical reactions $5Al_2O_3 + Ca_{12}Al_{14}O_{33} \rightarrow 12CaAl_2O_4$ ($\Delta G = -172$ kJ/mol) and $5CaAl_4O_7 + Ca_{12}Al_{14}O_{33} \rightarrow 17CaAl_2O_4$ ($\Delta G = -122$ kJ/mol) begin at these temperatures.

If $m$ = 1/2, the sintering activation energy $Q_{s1}$ calculated according to Young-Cutler model corresponds accurately to the sintering activation energy determined from the slope of the $\ln\{\ln[((\rho/\rho_{th})/(1-\rho/\rho_{th})]\} - T_m/T$ dependence (Fig. 10). At low sintering temperatures for $m$ = 1/2, the effective activation energy according to Young-Cutler was close to the one of volume diffusion of oxygen ions in alumina (636 kJ/mol ~ 33 $kT_m$) [58]. It should be noted also that the sintering activation energy at high temperatures was abnormally small [12–13 $kT_m$ ~ 230–250 kJ/mol] [70, 71]. The values of $Q_s$ obtained were much smaller than the activation energy for the diffusion of oxygen ions along the grain boundaries in alumina [19.7–20.5 $kT_m$ ~ 380–395 kJ/mol [70, 71]]. In our opinion, these results indicate that local melting of the CaO-$Al_2O_3$ phase may begin at high temperatures. The acceleration of the melting process may originate from the presence of Na and Si impurities in the alumina powder forming NaO and $SiO_x$ species, respectively. Local melting of the CaO-$Al_2O_3$ phase may provoke the anomalous grain growth in alumina.

## 4.2 Results of mechanical tests of the ceramics

At present, there are some works [15, 18, 72] demonstrating the potential to improve the mechanical properties of alumina by forming a bimodal grain size distribution. The bimodal grain size distribution in these ceramics was achieved by sintering a mixture of powders with different particle sizes. In particular, the work [18] demonstrated the possibility of increasing the ultimate strength of alumina ceramics by adding 20 and 30 wt.% of ultrafine α-$Al_2O_3$ powder and sintering at 1450 °C. The ceramics with 20 and 30 wt.% ultrafine α-$Al_2O_3$ powder added sintered at 1450 °C had the ultimate strengths 426 ± 52 MPa and 437 ± 28 MPa, respectively. The ceramics sintered at 1450 °C from fine grained α-$Al_2O_3$ powders (without the addition of the ultrafine powders) had the ultimate strength of 353 ± 37 MPa. However, no positive effect of the bimodal grain size distribution on the fracture toughness of alumina was observed.

In the present study, we controlled the parameters of the bimodal size distribution in alumina by adjusting the sintering conditions and adding 0.25% MgO.

The results of the mechanical tests of the ceramics using the B3B method are presented in Table 4. From the results shown in Table 4, one can see the ceramics A sintered in Regime I and Regime II to have the lowest strengths. The addition of 0.25% MgO allowed increasing the ultimate strength of alumina sintered in Regime I from 300 up to 520 MPa. The ultimate strength of alumina sintered in Regime II with the addition of 0.25% MgO increased from 255 up to 330 MPa that exceeds the dispersion of the ultimate strength (30-40 MPa) only slightly. The ultimate strength of ceramic B sintered in Regime III was 350-385 MPa.

Let us analyze the effect of the microstructure parameters of alumina on its mechanical properties.

As noted above, the positive effect of the bimodal grain size distribution on the mechanical properties of the ceramics may originate from the deviation of the crack propagation paths from the straight lines. This leads to an increased energy dissipation required for the crack growth and contributes to the enhancement of the strength of the ceramic. It is important to note that a higher volume fraction and/or larger sizes of the abnormal grains may lead to cracks with the intergranular characteristics, which can propagate rapidly within large grains resulting in worsening in worsening the mechanical properties of the ceramics. Fig. 13a shows the dependencies of the ultimate strength and of the microhardness of alumina on the volume fraction of the abnormally large grains. Fig. 13b illustrates the dependence of the mechanical properties of alumina on the form factor $F$ of the abnormally large grains. From Fig. 13, one can see the highest ultimate strength to be achieved in the ceramics with the volume fraction of the abnormally large grains of 12%, and low form factor ($F$ ~ 2).

It is worth noting that a significant reduction in the ultimate strength of alumina was observed even with a slight increase in the form factor $F$ (Fig. 13b). This is quite an unexpected result as the presence of the elongated grains in alumina leads to more significant deviation of the cracks from the straight-line trajectories. In our opinion, an anisotropy of the crystal lattice may be an additional factor contributing to the worsening of the mechanical properties of alumina with the bimodal grain size distribution and large values of the form factor $F$. The lattice anisotropy results in an anisotropy of the thermal expansion coefficient and in the formation of the tensile strain fields near the boundaries of the abnormally large grains [15, 17].

In materials with bimodal grain size distribution, the average grain sizes ($d_{av}$) can be conveniently calculated using a formula that takes into account the size and volume fraction of small and abnormally large grains:

$$d_{av} = D_{AGG} \cdot f_v + d(1 - f_v), \quad (2)$$

where $D_{AGG}$ and $f_v$ are the size and volume fraction of the abnormally large grains and $d$ is the size of the small grains. This approach is applied often to describe the abnormal grain growth during the annealing of severely deformed metals [73]. Knowing the effective area of the abnormally large grains $S_{AGG}$, one can calculate their effective size using the formula: $D_{AGG} = \sqrt{S_{AGG}/\pi}$. The dependencies of the ultimate strength and hardness on the average grain size $d_{av}$ for $F$ = 2-3 are presented in Fig. 14.

From Fig. 14, one can see the highest ultimate strength and hardness of alumina were achieved for the average grain size $d_{av}$ ~ 1.5 μm. These results agree well with the findings in [74-76] where the effect of average grain size on the dynamic strength and microhardness of fine-grained alumina was investigated. The results obtained indicate indirectly the correctness of the procedure for calculating the average grain size in the ceramics with the bimodal grain size distribution.

The ceramic with the average grain size $d_{av}$ ~ 1.5 μm was featured by a small volume fraction of abnormally large grains ($f_v$ = 12%) with the average size $D_{AGG}$ ~ 6 μm and the aspect ratio $F$ ~ 2. Increasing the volume fraction of the large grains up to 30% or increasing the average size of the abnormally large grains up to ≈8 μm led to drastic decrease in the ultimate strength of alumina.

In our opinion, the non-monotonic dependence of the ultimate strength of alumina on the average grain size can be attributed to the combination of two factors. On one hand, the presence of abnormally large grains increases the crack path length, as the crack has to pass by these large grains. The presence of CaO-$Al_2O_3$ or MgO ($MgAl_2O_4$) particles at the grain boundaries contributes to transcrystalline fracture behavior. On the other hand, the presence of large grains and increasing the average grain size leads to the generation of the tensile internal microstresses near the grain boundaries [9]. This results in reduced strength and hardness of alumina.

**Conclusion**

1. It has been shown that during sintering at constant heating rate (Regime I), the increase in the heating rate does not affect significantly the density and average grain sizes of the fine-grained matrix while there is a tendency to decrease in the volume fraction and sizes of the large grains. The use of MgO additive and the reduction in the compaction rate by applying multi-stage heating (Regime II) allows reducing the nonuniformity of the microstructure – the decreasing of the volume fraction and sizes of the abnormally large grains. The application of the multi-stage heating with subsequent isothermal holding at reduced temperatures (Regime III) allows controlling the kinetics of the bimodal grain size distribution formation.

2. The formation of a bimodal structure with abnormally large grains, the grain boundaries of which are characterized by lower strength than the grains themselves, enables an increase in the crack resistance of aluminum oxide by more than two times compared to ceramics sintered using the conventional continuous heating regime (Regime I). An increase in the volume fraction of abnormally large grains significantly influences the trajectory of crack propagation. In ceramics with a higher volume fraction of abnormally large grains, a significant deviation of the crack from a straight trajectory is observed.

3. The formation of the bimodal grain size distribution allows improving the hardness and strength of alumina when tested by b3b method. The formation of the optimal bimodal grain size distribution with a low volume fraction of the abnormally large grains ($f_v \sim 12\%$) with a low form factor ($F \sim 2$) is efficient for enhancing the strength of alumina. The effective average sizes of abnormally large grains $D_{AGG}$ were ≈ 6 µm, the average grain sizes in the fine-grained matrix d were 0.8-0.9 µm, and the average grain sizes in the ceramics $d_{av}$ were ~ 1.5 µm.

4. A decrease in the strength and in the hardness of the ceramics with increased grain form factor has been observed. The worsening of the mechanical properties of the ceramics with a high form factor was attributed to the formation of tensile internal micro-stresses near the boundaries of the abnormally large grains, which originate from the anisotropy of the crystal lattice of alumina leading to the anisotropy of the thermal expansion coefficient.

**Fig. 1** Possible crack propagation mechanisms in ceramics: (a) coarse grains with strong boundaries; (b) a bimodal structure with strong boundaries of large grains; and (c) a bimodal structure with weakened boundaries of large grains. The crack is highlighted in red, while the boundary of an abnormally large grain with reduced strength is marked in blue. Scheme.

**Fig. 2** Synthesis scheme of the α-$Al_2O_3$ powders

**Fig. 3** XRD curves of the precursor and of the precursor (a) and synthesized α-$Al_2O_3$ powder (b)

**Fig. 4** TG and DSC curves of the precursor powder: 1 – TG, 2 – DSC

**Fig. 5** SEM images of the initial powders: (a) precursor; (b) synthesized α-$Al_2O_3$ powder

**Fig. 6** Size distribution of the alumina agglomerates. DLS

**Fig. 7** Temperature curves of the shrinkage $L(T)$ and shrinkage rate $S(T)$ for powders A (a) and B (d) at the heating rates (2.5, 5, 10, and 20 °C·min$^{-1}$)

**Fig. 8** Microstructure of ceramics A (a) and B (b), sintered in Regime I. $V_h$ = 10 °C·min$^{-1}$. SEM

**Fig. 9** Comparison of the *L(T)* (dashed line) and $S(T)$ (solid line) curves for ceramics A (black line) and B (red line). Sintering in Regime II

**Fig. 10** Microstructure of ceramics A (a) and B (b) sintered in Regime II. SEM

**Fig. 11** SEM images of microstructure of ceramics obtained by heating to $T_1$ = 1550°C with a 3-hour holding at $T_2$ = 1300 (a), 1350 (b), 1400 (c), 1450 (d), and 1500 °C (e). The grain size distribution histogram in the ceramic sintered at $T_2$ = 1450 °C (f)

**Fig. 12** Determining of the sintering activation energy. Dependencies of the powder shrinkage of the α-$Al_2O_3$ (powder A) on the heating temperature in the $\ln(T\cdot\partial\varepsilon/\partial T) - T_m/T$ axes (model [56, 57]) and in the $\ln\{\ln[(\rho/\rho_{th})/(1-\rho/\rho_{th})]\} - T_m/T$ (model [65])

**Fig. 13** Effect of the volume fraction of the abnormally large grains $f_v$ (a) and of their form factor $F$ (b) on the mechanical properties of the alumina. Blue circles – ultimate strength, red squares – hardness

**Fig. 14** Dependencies of the ultimate strength (blue circles) and of the hardness (red squared) of alumina on the average grain size

**Table 1**. Summary of the dilatometric investigations results on the sintering kinetics of the $Al_2O_3$ (powder A) and $Al_2O_3$ + 0.25%MgO (powder B)

| $V_h$, °C·min$^{-1}$ | Powder A | | | | Powder B | | | |
|---|---|---|---|---|---|---|---|---|
| | $T_{max(1)}$, °C | $S_{max(1)}$, %·s$^{-1}$ | $T_{max(2)}$, °C | $S_{max(2)}$, %·s$^{-1}$ | $T_{max(1)}$, °C | $S_{max(1)}$, %·s$^{-1}$ | $T_{max(2)}$, °C | $S_{max(2)}$, %·s$^{-1}$ |
| 2.5 | 1370 | $0.33\cdot10^{-5}$ | 1510 | $0.45\cdot10^{-5}$ | 1370 | $0.30\cdot10^{-5}$ | 1480 | $0.67\cdot10^{-5}$ |
| 5 | 1395 | $0.77\cdot10^{-5}$ | 1520 | $0.81\cdot10^{-5}$ | 1390 | $0.67\cdot10^{-5}$ | 1495 | $1.1\cdot10^{-5}$ |
| 10 | 1420 | $1.3\cdot10^{-5}$ | 1530 | $1.7\cdot10^{-5}$ | 1405 | $1.3\cdot10^{-5}$ | 1510 | $2.0\cdot10^{-5}$ |
| 20 | 1440 | $2.5\cdot10^{-5}$ | 1540 | $3.4\cdot10^{-5}$ | 1440 | $2.4\cdot10^{-5}$ | 1510 | $4.1\cdot10^{-5}$ |

**Table 2**. Characteristics of sintered ceramics obtained in Regime I

| Ceramics | $V_h$, °C·min$^{-1}$ | $\rho/\rho_{th}$, % | *d*, μm | Parameters of large grains | | | | | Hv, GPa | $K_{IC}$, МПа·м$^{1/2}$ |
|---|---|---|---|---|---|---|---|---|---|---|
| | | | | $f_v$, % | $L_{AGG}$, μm | $l_{AGG}$, μm | *F* | $S_{AGG}$, μm$^2$ | | |
| *A* | 2.5 | 96.8 ± 1.2 | 0.9 ± 0.1 | 90 | 45 ± 13 | 10 ± 4 | 4.4 | 464 | 14.0 ± 1.7 | 2.6 |
| | 5 | 95.5 ± 1.4 | 0.8 ± 0.1 | 68 | 45 ± 10 | 9 ± 3 | 4.8 | 419 | 13.6 ± 1.1 | 2.3 |
| | 10 | 96.8 ± 1.2 | 0.8 ± 0.1 | 51 | 30 ± 10 | 8 ± 2 | 3.6 | 249 | 13.3 ± 1.1 | 2.0 |
| | 20 | 96.3 ± 1.3 | 0.9 ± 0.1 | 96 | - | - | - | - | 13.1 ± 0.9 | 2.2 |
| *B* | 2.5 | 98.0 ± 0.9 | 1.0 ± 0.1 | 87 | 15 ± 6 | 9 ± 3 | 1.7 | 135 | 16.2 ± 1.4 | - |
| | 5 | 97.5 ± 1.2 | 0.9 ± 0.1 | 45 | 17 ± 9 | 8.5 ± 3 | 2.0 | 142 | 16.8 ± 1.6 | 4.9 |
| | 10 | 97.7 ± 0.2 | 0.9 ± 0.1 | 12 | 16 ± 6 | 8 ± 3 | 2.0 | 123 | 19.5 ± 1.9 | 4.6 |
| | 20 | 99.1 ± 1.2 | 0.9 ± 0.1 | 12 | 13 ± 5 | 7 ± 3 | 1.9 | 81 | 18.2 ± 0.9 | 4.5 |

**Table 3**. Characteristics of ceramic B sintered in Regime III

| $T_2$, °C | $\rho/\rho_{th}$, % | *d*, µm | Parameters of large grains | | | | | | | |
|---|---|---|---|---|---|---|---|---|---|---|
| | | | $f_v$, % | $L_{AGG}$, µm | $l_{AGG}$, µm | *F* | $S_{AGG}$, µm² | Hv, GPa | $K_{IC}$, MPa·m$^{1/2}$ | $\sigma_b$, MPa |
| 1300 | 99.8 ± 1.2 | 0.7 ± 0.1 | 2 | 27 ± 6 | 9.5 ± 1 | 2.8 | 252 | 18.0 ± 1.0 | 4.6 | - |
| 1350 | 96.2 ± 0.2 | 0.7 ± 0.1 | 5 | 13 ± 5 | 5 ± 2 | 2.6 | 65 | 19.0 ± 4.1 | 5.4 | 350 ± 40 |
| 1400 | 96.1 ± 2.3 | 0.7 ± 0.1 | 30 | 23 ± 8 | 10 ± 4 | 2.4 | 219 | 18.4 ± 2.8 | 3.7 | 385 ± 45 |
| 1450 | 97.5 ± 1.9 | 0.7 ± 0.1 | 60 | 35 ± 11 | 14 ± 9 | 2.5 | 490 | 16.7 ± 2.3 | 4.5 | 300 ± 35 |
| 1500 | 99.5 ± 1.2 | 0.8 ± 0.1 | 78 | 55 ± 15 | 29 ± 6 | 1.9 | 1595 | - | - | 290 ± 50 |

**Table 4**. Results of mechanical tests of the ceramics using the B3B method

| Ceramic | Regime | $V_h$, °C·min$^{-1}$ | $T_1$, °C | $T_2$, °C | Microstructure parameters | | | | | $\sigma_b$, MPa |
|---|---|---|---|---|---|---|---|---|---|---|
| | | | | | $f_v$, % | *F* | $S_{AGG}$, µm² | $D_{AGG}$, µm | $d_{av}$, µm | |
| A | I | 10 | 1650 | - | 51 | 3.6 | 249 | 8.9 | 4.9 | 300 ± 40 |
| | II | - | 1565 | - | 5 | 5.8 | 145 | 6.8 | 1.2 | 255 ± 30 |
| B | I | 10 | 1650 | - | 12 | 2.0 | 123 | 6.2 | 1.5 | 520 ± 50 |
| | II | - | 1565 | - | 5 | 2.4 | 152 | 7.0 | 1.0 | 330 ± 40 |
| | III | - | 1550 | 1350 | 5 | 2.6 | 65 | 4.5 | 1.0 | 350 ± 40 |
| | III | - | 1550 | 1400 | 30 | 2.4 | 219 | 8.4 | 3.1 | 385 ± 45 |
| | III | - | 1550 | 1450 | 60 | 2.5 | 490 | 12.5 | 7.9 | 350 ± 45 |

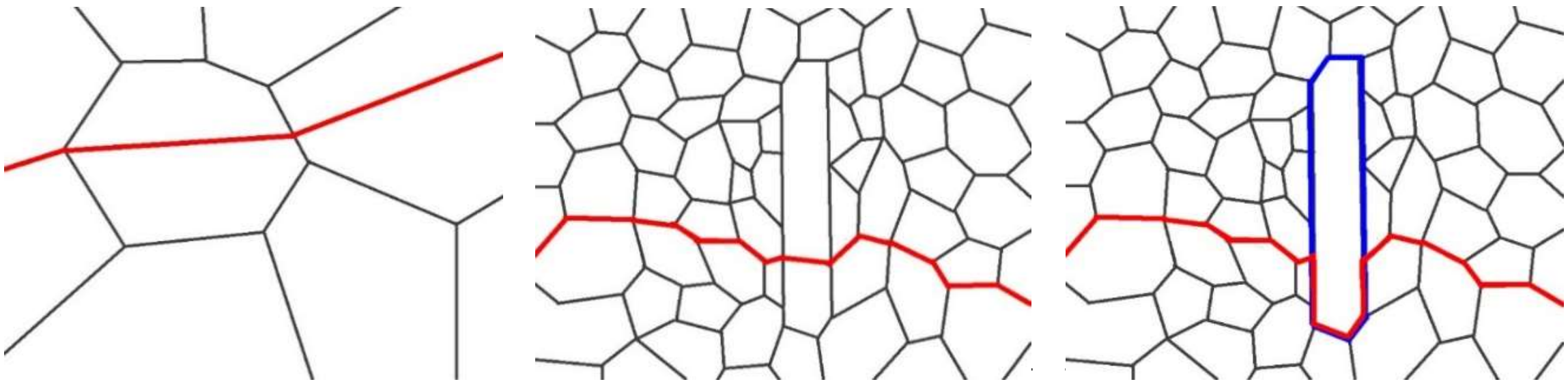

(a) (b) (c)

**Fig. 1**

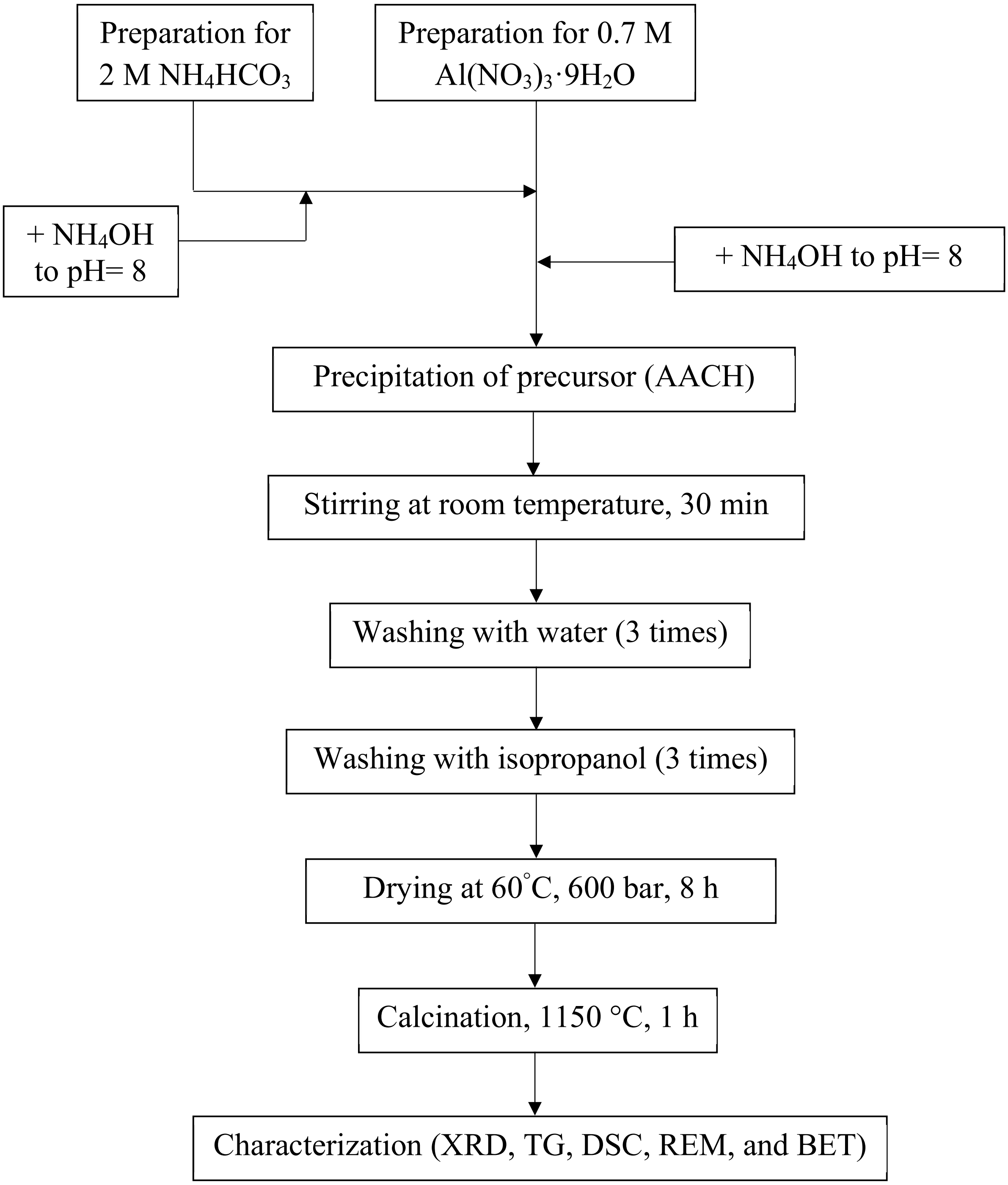
Preparation for 2 M $NH_4HCO_3$
Preparation for 0.7 M $Al(NO_3)_3 \cdot 9H_2O$
+ $NH_4OH$ to pH= 8
+ $NH_4OH$ to pH= 8
Precipitation of precursor (AACH)
Stirring at room temperature, 30 min
Washing with water (3 times)
Washing with isopropanol (3 times)
Drying at 60°C, 600 bar, 8 h
Calcination, 1150 °C, 1 h
Characterization (XRD, TG, DSC, REM, and BET)

**Fig. 2**

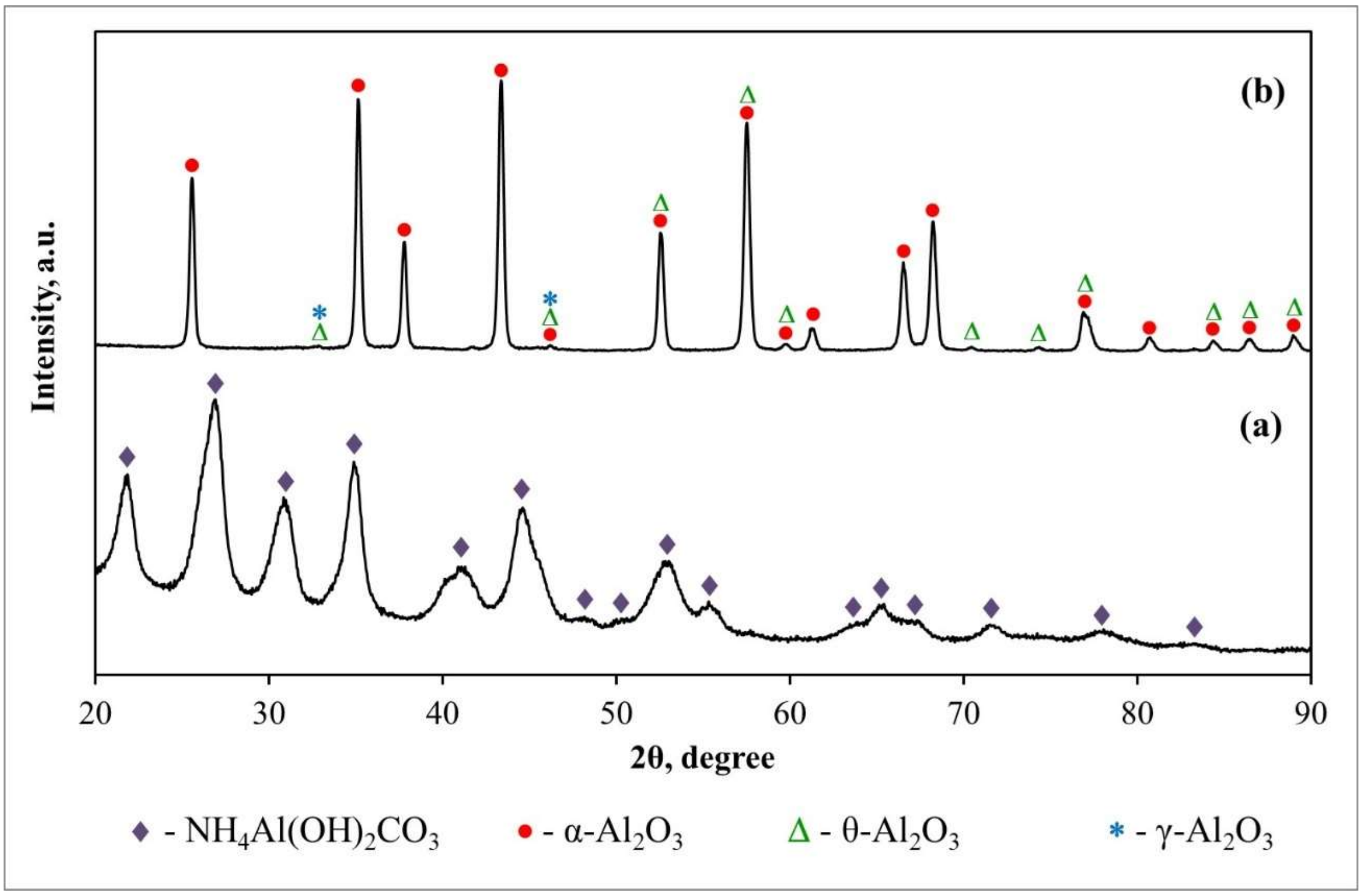

**Fig. 3**

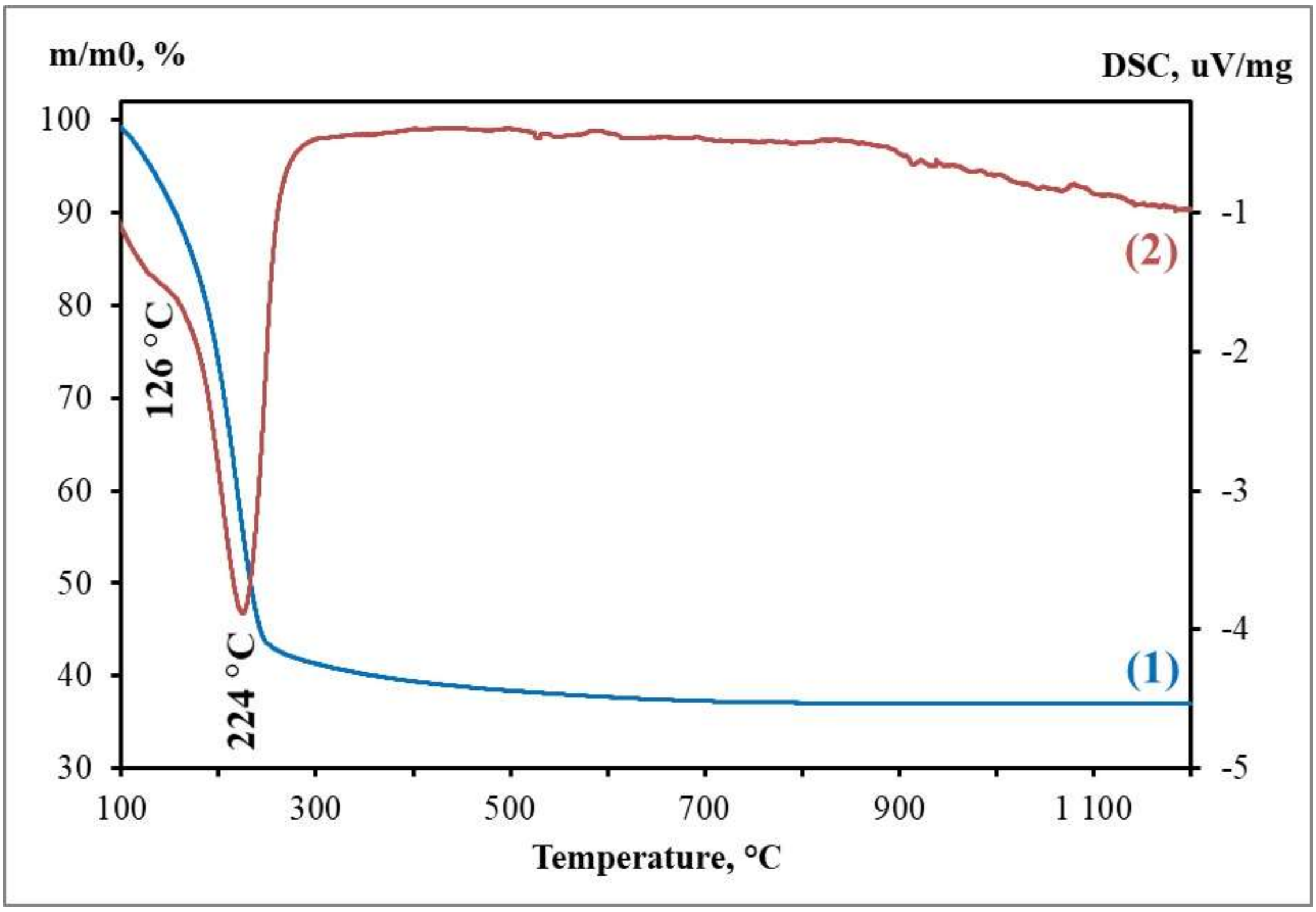

**Fig. 4**

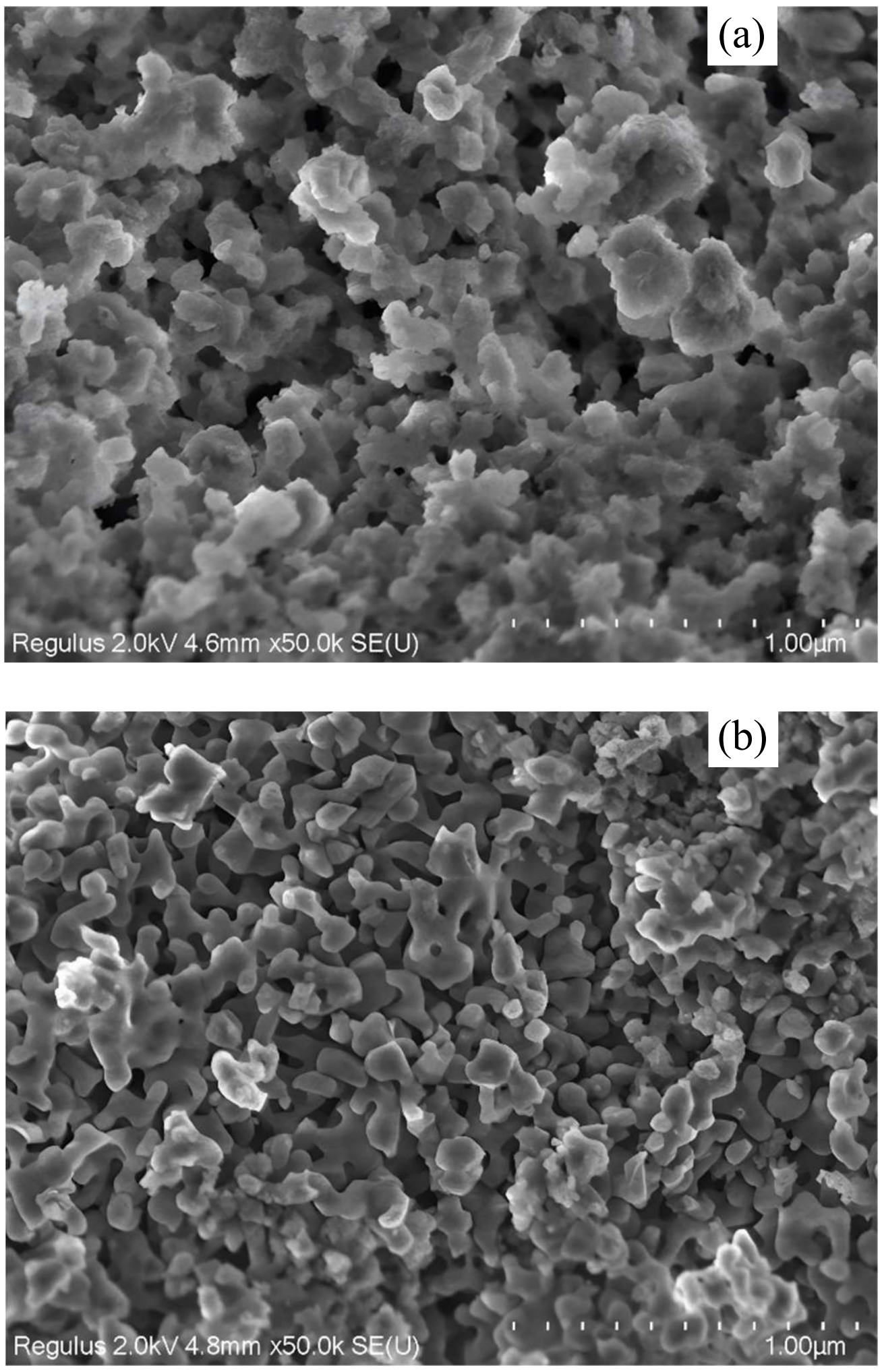
(a)
Regulus 2.0kV 4.6mm x50.0k SE(U)
1.00µm
(b)
Regulus 2.0kV 4.8mm x50.0k SE(U)
1.00µm

**Fig. 5**.

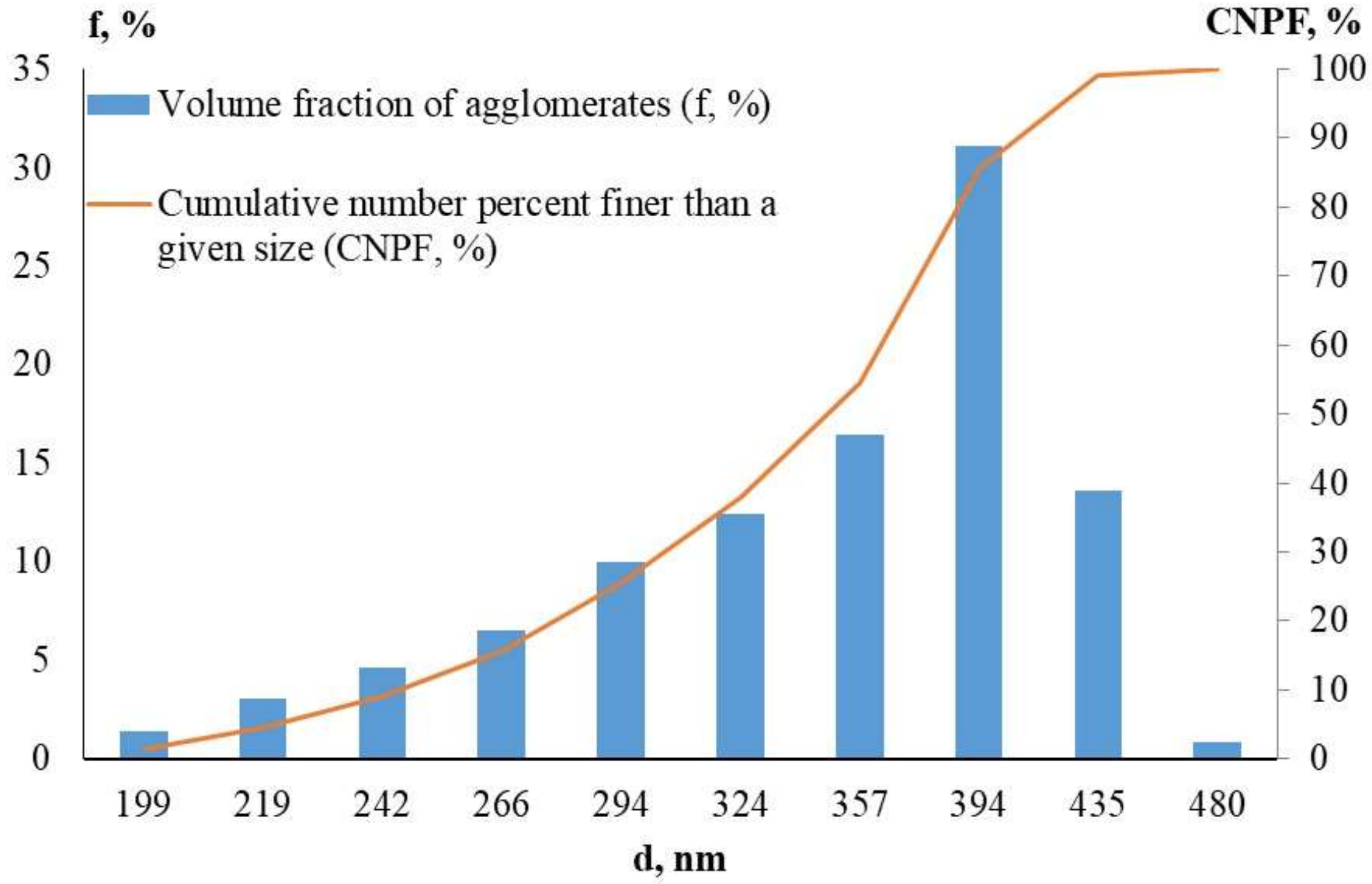

f, %
CNPF, %
35
30
25
20
15
10
5
0
100
90
80
70
60
50
40
30
20
10
0
Volume fraction of agglomerates (f, %)
Cumulative number percent finer than a given size (CNPF, %)
199
219
242
266
294
324
357
394
435
480
d, nm

**Fig. 6**.

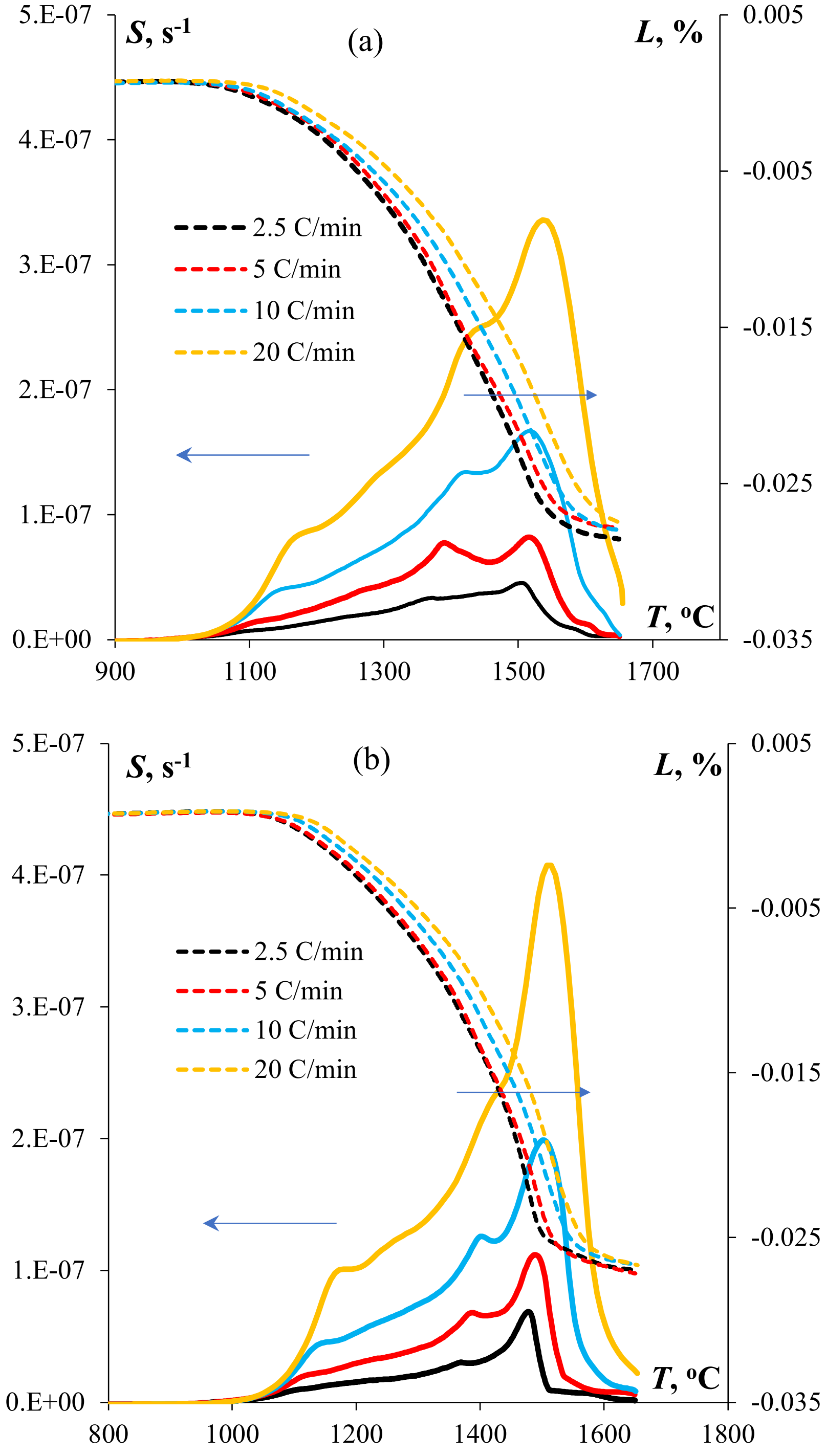

**Fig. 7**

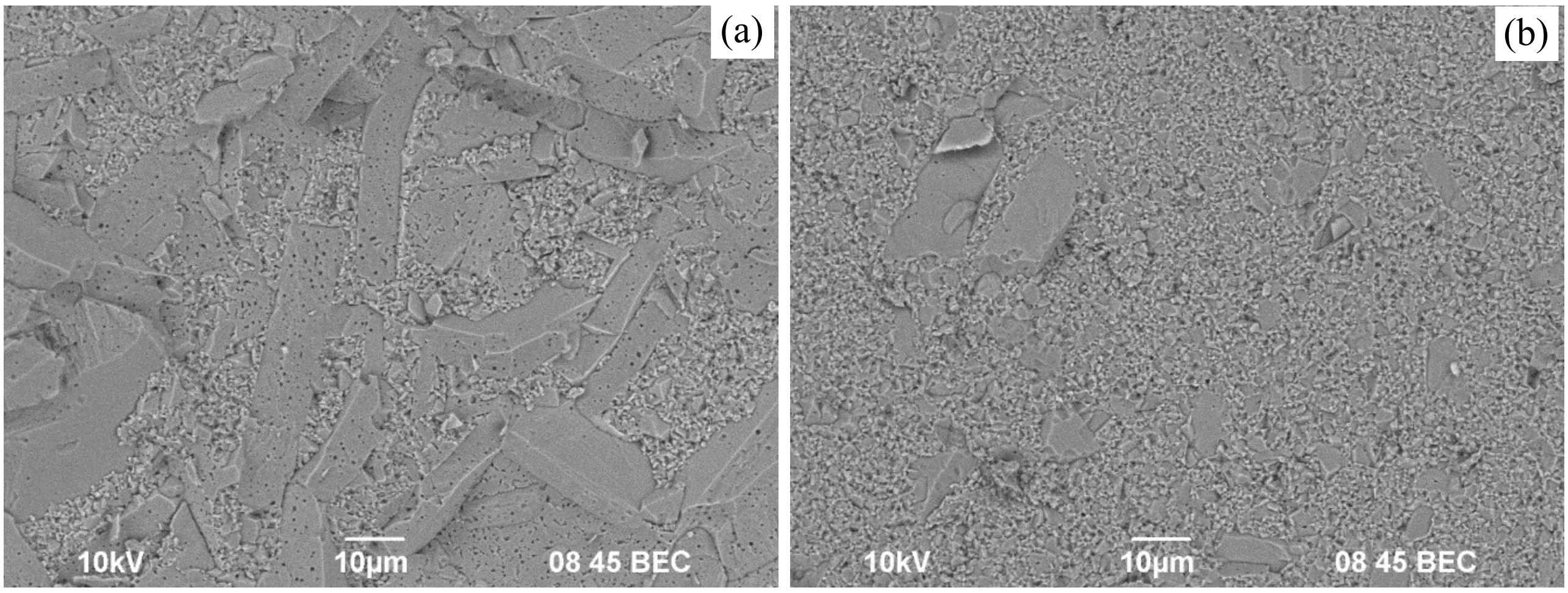

**Fig. 8**

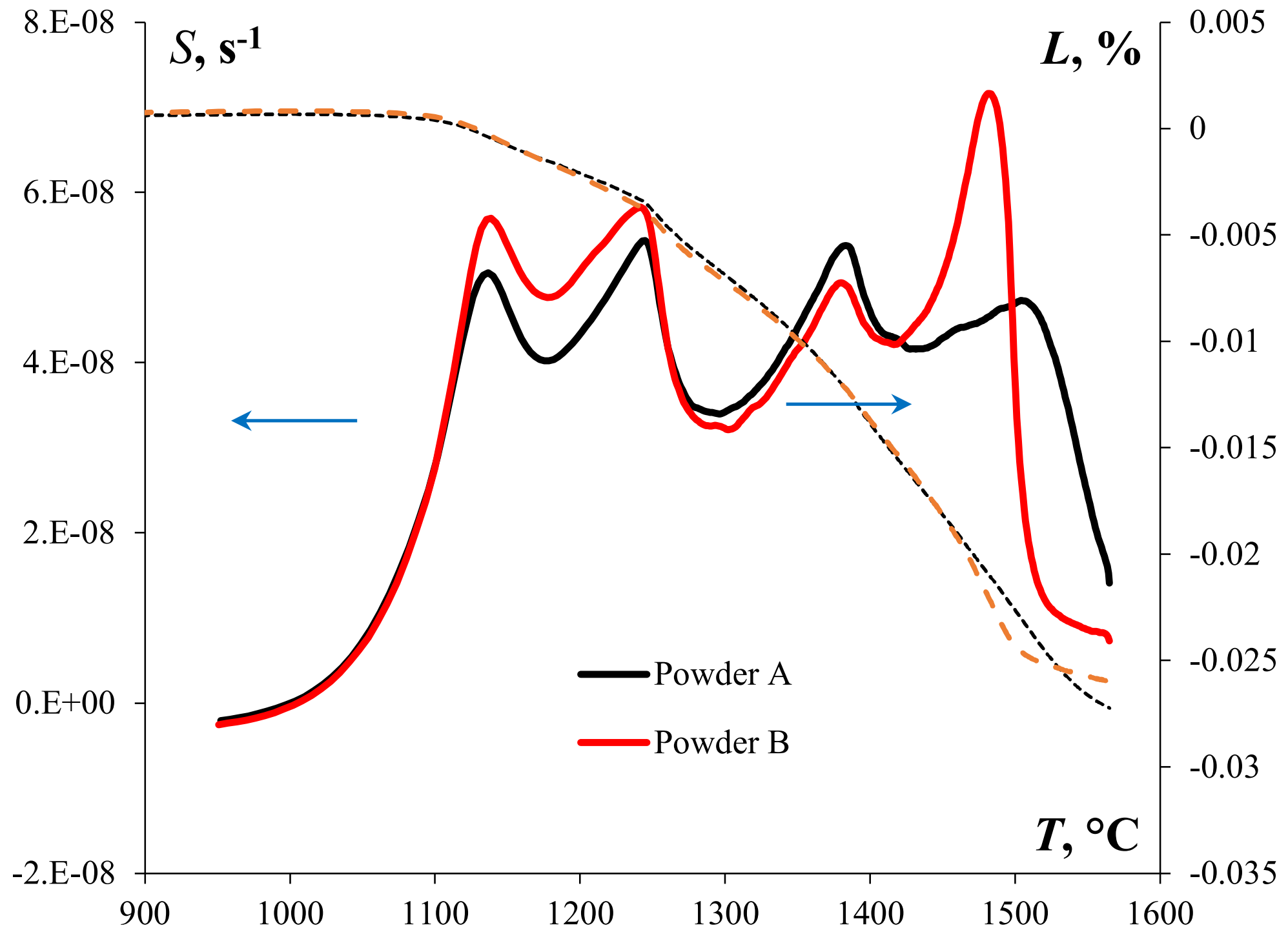

**Fig. 9**

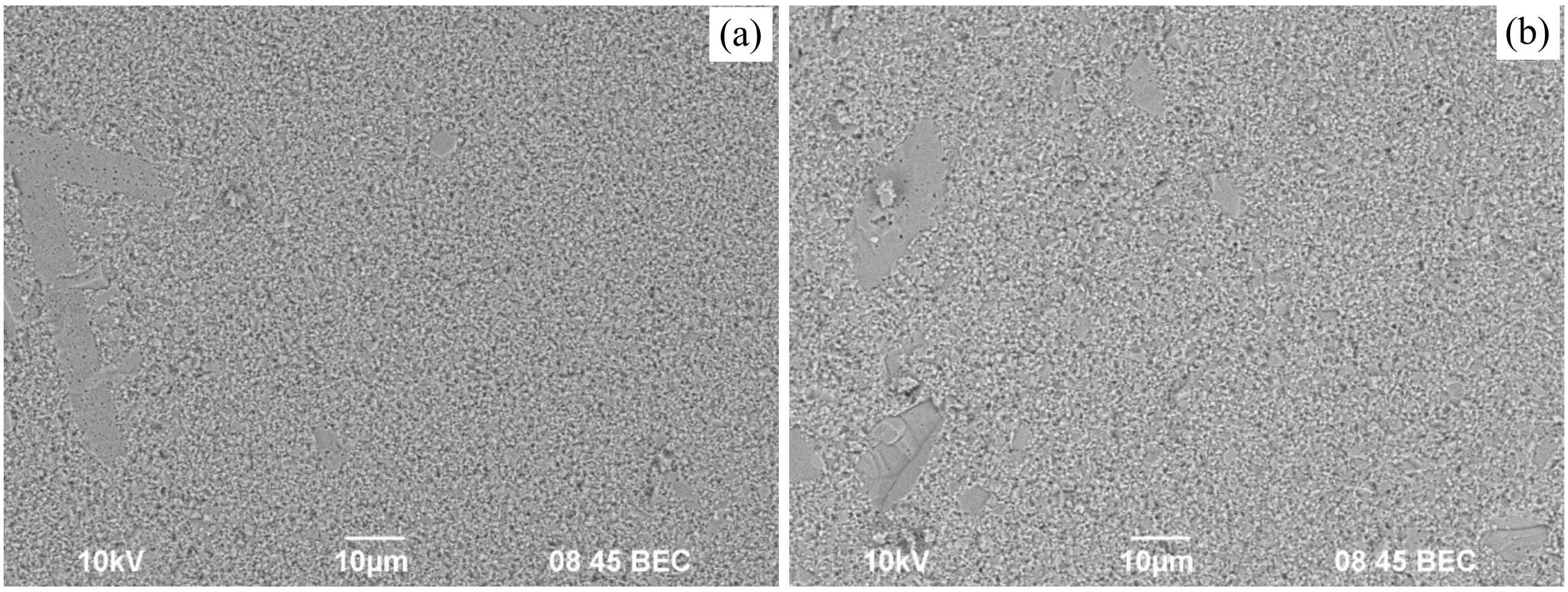

**Fig. 10**

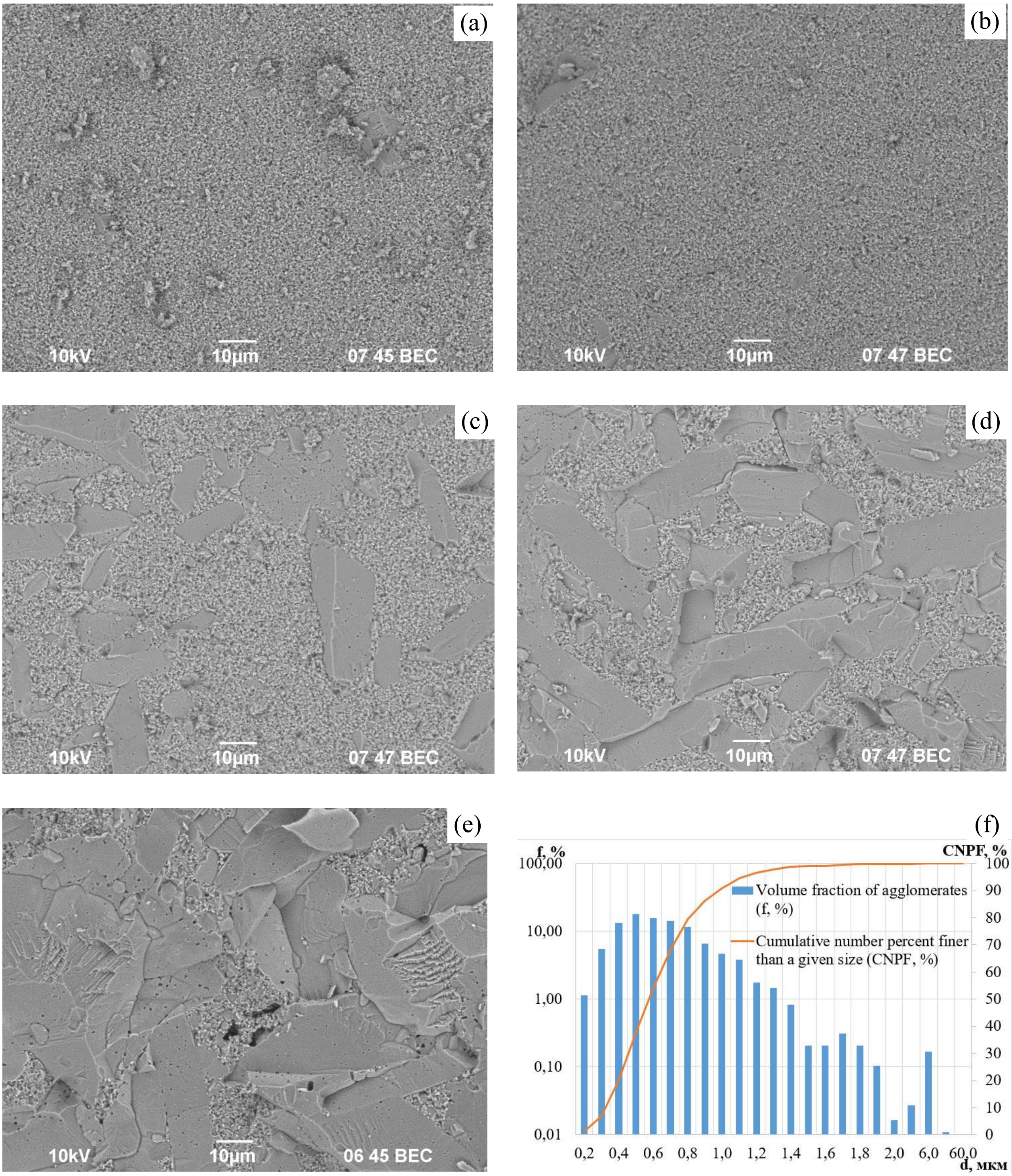
(a)
10kV
10µm
07 45 BEC
(b)
10kV
10µm
07 47 BEC
(c)
10kV
10µm
07 47 BEC
(d)
10kV
10µm
07 47 BEC
(e)
10kV
10µm
06 45 BEC
(f)
f, %
CNPF, %
Volume fraction of agglomerates (f, %)
Cumulative number percent finer than a given size (CNPF, %)
100,00
10,00
1,00
0,10
0,01
100
90
80
70
60
50
40
30
20
10
0
0,2
0,4
0,6
0,8
1,0
1,2
1,4
1,6
1,8
2,0
6,0
60,0
d, мкм

**Fig. 11.**

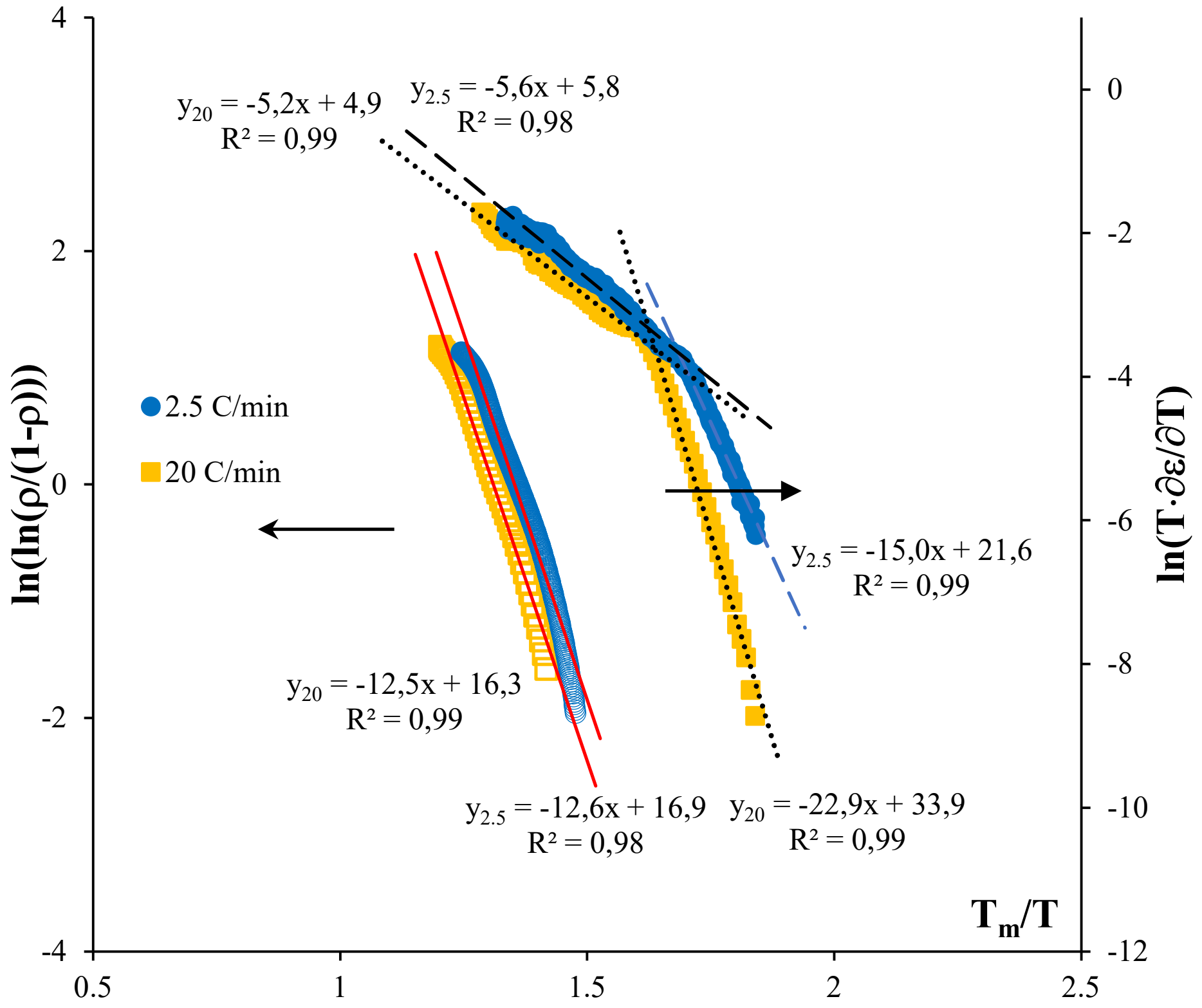

**Fig. 12**

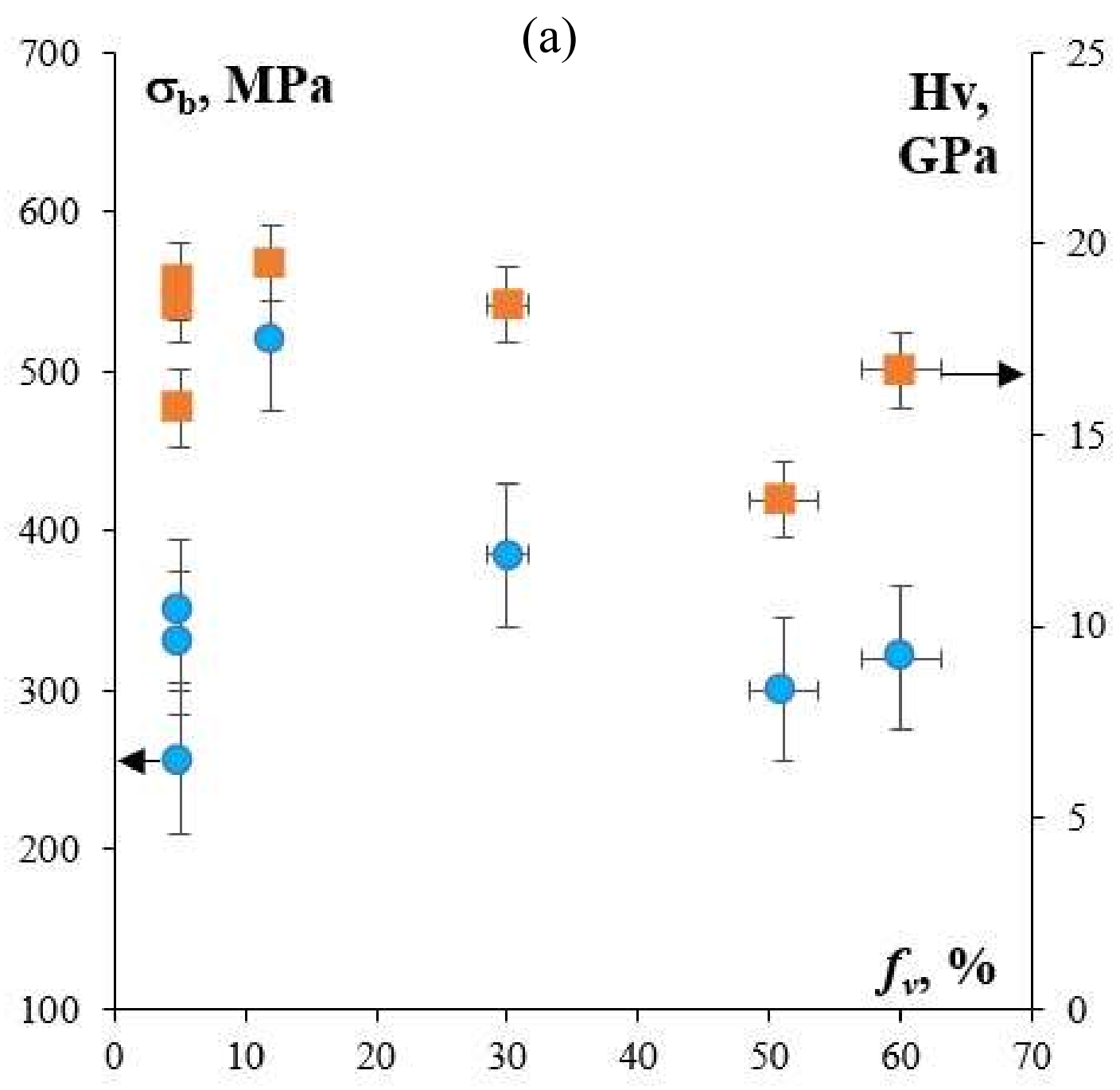

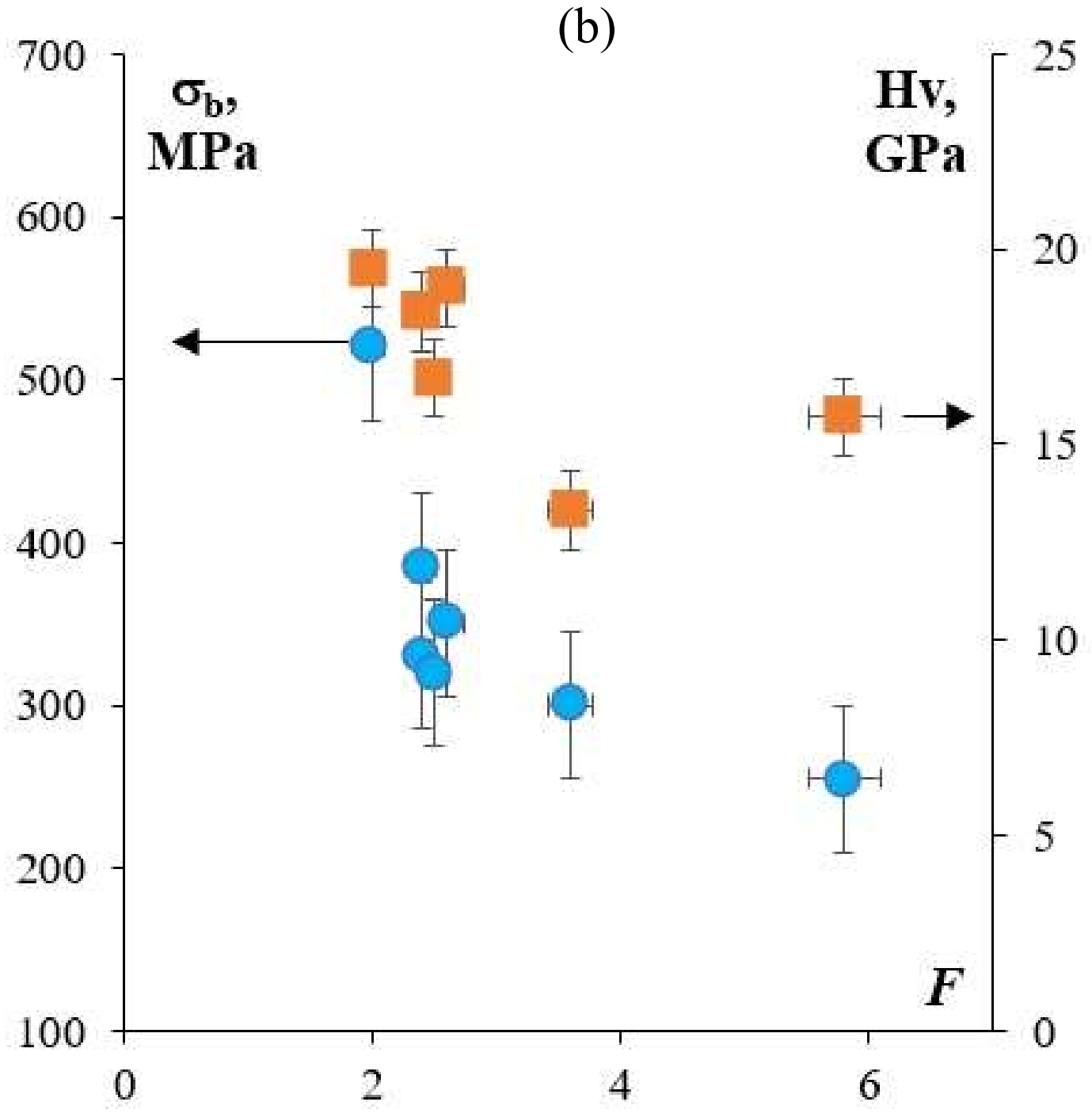

**Fig. 13**.

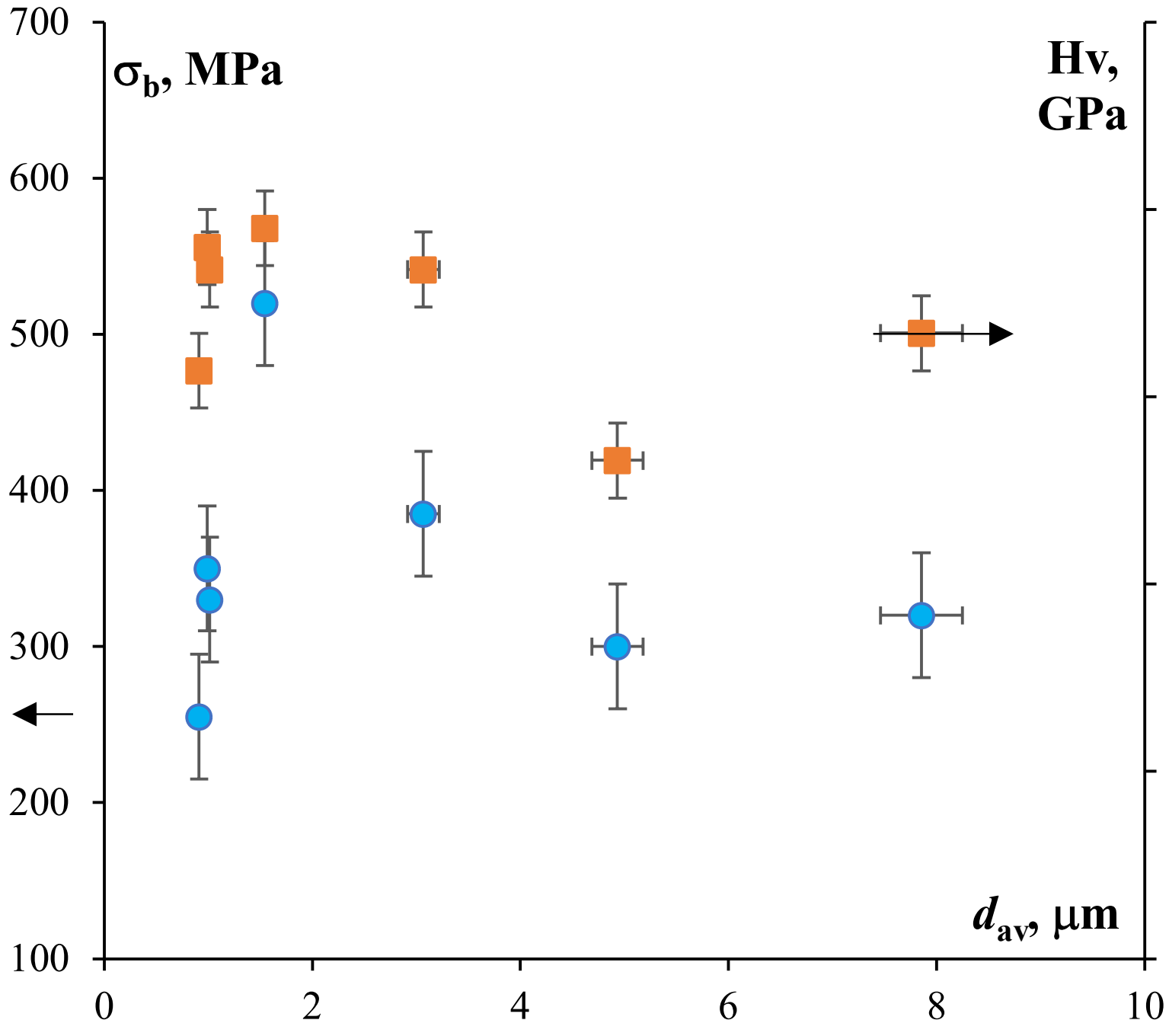

**Fig. 14**.